\documentclass[10pt,journal,compsoc]{IEEEtran}

\usepackage{graphicx}
\usepackage{textcomp}
\usepackage{array}
\usepackage{epstopdf}
\usepackage{subfigure}
\usepackage{caption}
\usepackage{multirow}
\usepackage{titlesec}
\usepackage{setspace}
\usepackage{algorithm}
\usepackage{algorithmic}
\usepackage[misc]{ifsym}
\usepackage{color}
\usepackage{framed} 

\usepackage{url}

\newcommand{\ie}{\textit{i.e.,} }
\newcommand{\eg}{\textit{e.g.,} }

\newcommand\opt[1]{}
\newcommand\find[1]{}

\newcommand{\ls}[1]
   {\dimen0=\fontdimen6\the\font 
    \lineskip=#1\dimen0
    \advance\lineskip.5\fontdimen5\the\font
    \advance\lineskip-\dimen0
    \lineskiplimit=.9\lineskip
    \baselineskip=\lineskip
    \advance\baselineskip\dimen0
    \normallineskip\lineskip
    \normallineskiplimit\lineskiplimit
    \normalbaselineskip\baselineskip
    \ignorespaces
   }

\newenvironment{smalldescription}{
   \setlength{\topsep}{0pt}
   \setlength{\partopsep}{0pt}
   \setlength{\parskip}{0pt}
   \begin{description}
   \setlength{\leftmargin}{.2in}
   \setlength{\parsep}{0pt}
   \setlength{\parskip}{0pt}
   \setlength{\itemsep}{0pt}}{\end{description}}

\begin{document}

\title{Contrastive Learning for Robust Android Malware Familial Classification}

\author{Yueming Wu,
        Shihan Dou,
        Deqing Zou,
        Wei Yang,
        Weizhong Qiang,
        and Hai Jin,~\IEEEmembership{Fellow, IEEE}
\IEEEcompsocitemizethanks{
\IEEEcompsocthanksitem Y. Wu, D. Zou (corresponding author), and W. Qiang are with National Engineering Research Center for Big Data Technology and System, Services Computing Technology and System Lab, Hubei Engineering Research Center on Big Data Security, School of Cyber Science and Engineering, Huazhong University of Science and Technology, Wuhan, 430074, China.
E-mail: wuyueming21@gmail.com, deqingzou@hust.edu.cn, wzqiang@hust.edu.cn
\IEEEcompsocthanksitem S. Dou is with Shanghai Key Laboratory of Intelligent Information Processing, School of Computer Science,
Fudan University, Shanghai, 200433, China.
E-mail: shihandou@foxmail.com
\IEEEcompsocthanksitem W. Yang is with University of Texas at Dallas, Dallas, USA.
E-mail: wei.yang@utdallas.edu
\IEEEcompsocthanksitem H. Jin is with National Engineering Research Center for Big Data Technology and System, Services Computing Technology and System Lab, Cluster and Grid Computing Lab, School of Computer Science and Technology, Huazhong University of Science and Technology, Wuhan, 430074, China.
E-mail: hjin@hust.edu.cn
}
}

\IEEEtitleabstractindextext{
\begin{abstract}
Due to its open-source nature, Android operating system has been the main target of attackers to exploit.
Malware creators always perform different code obfuscations on their apps to hide malicious activities.
Features extracted from these obfuscated samples through program analysis contain many useless and disguised features, which leads to many false negatives.
To address the issue, in this paper, we demonstrate that obfuscation-resilient malware family analysis can be achieved through contrastive learning.
The key insight behind our analysis is that contrastive learning can be used to reduce the difference introduced by obfuscation while amplifying the difference between malware and other types of malware.
Based on the proposed analysis, we design a system that can achieve robust and interpretable classification of Android malware.
To achieve robust classification, we perform contrastive learning on malware samples to learn an encoder that can automatically extract robust features from malware samples.
To achieve interpretable classification, we transform the function call graph of a sample into an image by centrality analysis.
Then the corresponding heatmaps can be obtained by visualization techniques.
These heatmaps can help users understand why the malware is classified as this family.
We implement \emph{IFDroid} and perform extensive evaluations on two datasets.
Experimental results show that \emph{IFDroid} is superior to state-of-the-art Android malware familial classification systems.
Moreover, \emph{IFDroid} is capable of maintaining a 98.4\% F1 on classifying 69,421 obfuscated malware samples.
\end{abstract}

}

\maketitle

\IEEEdisplaynontitleabstractindextext

\IEEEpeerreviewmaketitle

\section{Introduction}

\IEEEPARstart{A}{s} the most widely used mobile operating system~\cite{report1}, the security of Android platform has become more and more closely related to personal privacy and financial security.
Meanwhile, due to the open-source and market openness of Android operating system, it is more likely to be exploited by malware~\cite{report2}.
To hide their malicious tasks, different code obfuscations have been applied by attackers~\cite{2018obfuscate1, 2018obfuscate2}.
After obfuscations, malware samples become more complex, resulting in features obtained from them containing many useless and camouflage features.
These futile features make it difficult to perform accurate behavioral analysis of Android malware.
Therefore, it is important to provide obfuscation-resilient Android malware analysis. 

Most traditional Android malware analysis methods~\cite{peng2012using,wang2014exploring,arp2014drebin} cannot resist code obfuscations.
For example, for familial classification of Android malware, it can be roughly divided into two main categories~\cite{fan2018android}, namely string-based approaches (\eg permissions~\cite{peng2012using}) and graph-based techniques (\eg function call graph~\cite{fan2019graph}).
Some methods~\cite{peng2012using,wang2014exploring,arp2014drebin} focus on permissions requested by apps and search for the presence of several strings (\eg API calls) from disassembling code to build models to analyze Android malware.
However, they can be easily evaded by obfuscations because of the lack of structural and contextual information of the program behaviors.
To achieve more robust malware classification, studies~\cite{fan2018android, fan2019graph} distill the program semantics of apps into graph representations and apply graph matching to analyze the malware families.
For example, \emph{FalDroid}~\cite{fan2018android} extracts the function call graph of an app and applies frequent subgraph analysis to classify Android malware.
However, Hammad and Dong \emph{et al.} \cite{2018obfuscate1, 2018obfuscate2} report that malware creators always perform complex obfuscations (\eg control-flow obfuscations) to hidden their malicious tasks.
In this case, features extracted from graphs obtained by \emph{FalDroid}~\cite{fan2018android} may not be accurate since graphs may change a lot after applying advanced code obfuscations.
In one word, due to different code obfuscations, features obtained from malware samples may contain many useless and disguised features, making it difficult to achieve accurate behavioral analysis.

To address the issue, we propose to use contrastive learning on Android malware analysis.
Due to the powerful high-level feature extraction of contrastive learning, it has been widely used in different areas, such as text representation learning~\cite{2020DeCLUTR} and language understanding~\cite{2020Cert}.
To the best of our knowledge, we are the first to use contrastive learning to resist code obfuscations.
To demonstrate the ability of contrastive learning on analyzing obfuscated Android malware, in this paper, we propose a novel approach that can achieve obfuscation-resilient Android malware classification.

Specifically, we first obtain the function call graph of an app and then apply centrality analysis~\cite{wu2019malscan} to transform the graph into an image.
The generated images are used to train an encoder by contrastive learning.
The use of contrastive learning is to maximize the similarity between positive samples and minimize the similarity between negative samples.
In practice, although applying obfuscations may change the app codes, the inherent program semantics do not change.
In other words, the obfuscated app can be treated as one of the positive samples of the original app.
Therefore, we can leverage contrastive learning to reduce the differences introduced by code obfuscations while enlarging the differences between different types of malware, making it possible to correctly classify the obfuscated malware into the corresponding family.

To further show how contrastive learning improves the usability of malware analysis, we apply visualization techniques to visualize the valuable features extracted by contrastive learning.
Specifically, we apply \emph{Gradient-weighted Class Activation Mapping++} (Grad-CAM++)~\cite{selvaraju2017gradcam, chattopadhay2018gradcam} on our images to obtain the corresponding heatmaps.
Grad-CAM++ is a class-discriminative localization technique that generates visual explanations for any CNN-based network without changing the architecture or retraining.
According to the intensity of the color in the heatmap, we can know which features are more effective in classifying this malware as this family.
These valuable features can represent the essential behaviors to explain why the malware is classified as this family.

We implement \emph{IFDroid} and conduct evaluations on two datasets.
Through the comparative experimental results, we find that \emph{IFDroid} is superior to ten state-of-the-art Android malware familial classification systems (\ie \emph{Dendroid}~\cite{suarez2014dendroid}, \emph{Apposcopy}~\cite{feng2014apposcopy}, \emph{DroidSIFT}~\cite{zhang2014droidsift}, \emph{MudFlow}~\cite{avdiienko2015mudflow}, \emph{DroidLegacy}~\cite{deshotels2014droidlegacy}, \emph{Astroid}~\cite{feng2016astroid}, \emph{FalDroid}~\cite{fan2018android}, \emph{AOM}~\cite{blanc2019identifyingAOM}, \emph{MVIIDroid}~\cite{wu2020mviidroidMVI}, and \emph{CDFG}~\cite{zhiwu2019androidCDFG}).
As for obfuscations, \emph{IFDroid} can maintain a 98.4\% F1 on classifying 69,421 obfuscated malware samples.
As for interpretability, our experiments show that the heatmaps of most malware in the same family are similar, and the heatmaps of malware in different families are different.
This result is in line with expectations and it can help security analysts analyze the specific reasons why malware samples are classified as corresponding families.
As for runtime overhead, \emph{IFDroid} requires an average of 1.78 seconds to complete the classification and 1.62 seconds to interpret the classification result in our dataset.
Such results indicate that \emph{IFDroid} can conduct large-scale malware analysis.

\par In summary, this paper makes the following contributions:
\begin{itemize}
\item{
To the best of our knowledge, we are the first to use contrastive learning to resist code obfuscations of Android malware.
Contrastive learning can not only improve the accuracy but also enhance the robustness of Android malware analysis.
}
\item{
We design a novel system (\ie \emph{IFDroid}) by transforming the function call graph of an app into an image and performing contrastive learning on generated images.
\emph{IFDroid} can achieve robust and interpretable classification of Android malware.
}
\item{
We conduct evaluations on two datasets and results indicate that \emph{IFDroid} is superior to ten state-of-the-art Android malware classification systems (\ie \emph{Dendroid}~\cite{suarez2014dendroid}, \emph{Apposcopy}~\cite{feng2014apposcopy}, \emph{DroidSIFT}~\cite{zhang2014droidsift}, \emph{MudFlow}~\cite{avdiienko2015mudflow}, \emph{DroidLegacy}~\cite{deshotels2014droidlegacy}, \emph{Astroid}~\cite{feng2016astroid}, \emph{FalDroid}~\cite{fan2018android}, \emph{AOM}~\cite{blanc2019identifyingAOM}, \emph{MVIIDroid}~\cite{wu2020mviidroidMVI}, and \emph{CDFG}~\cite{zhiwu2019androidCDFG}).
}
\end{itemize}

\par \noindent \textbf{Paper organization.} The remainder of the paper is organized as follows. 
Section 2 presents our motivation.  
Section 3 introduces our system. 
Section 4 reports the experimental results. 
Section 5 discusses the future work and limitations. 
Section 6 describes the related work. Section 7 concludes the present paper.

\section{Motivation Scenario}

\begin{figure}[htbp]
\centerline{\includegraphics[width=0.485\textwidth]{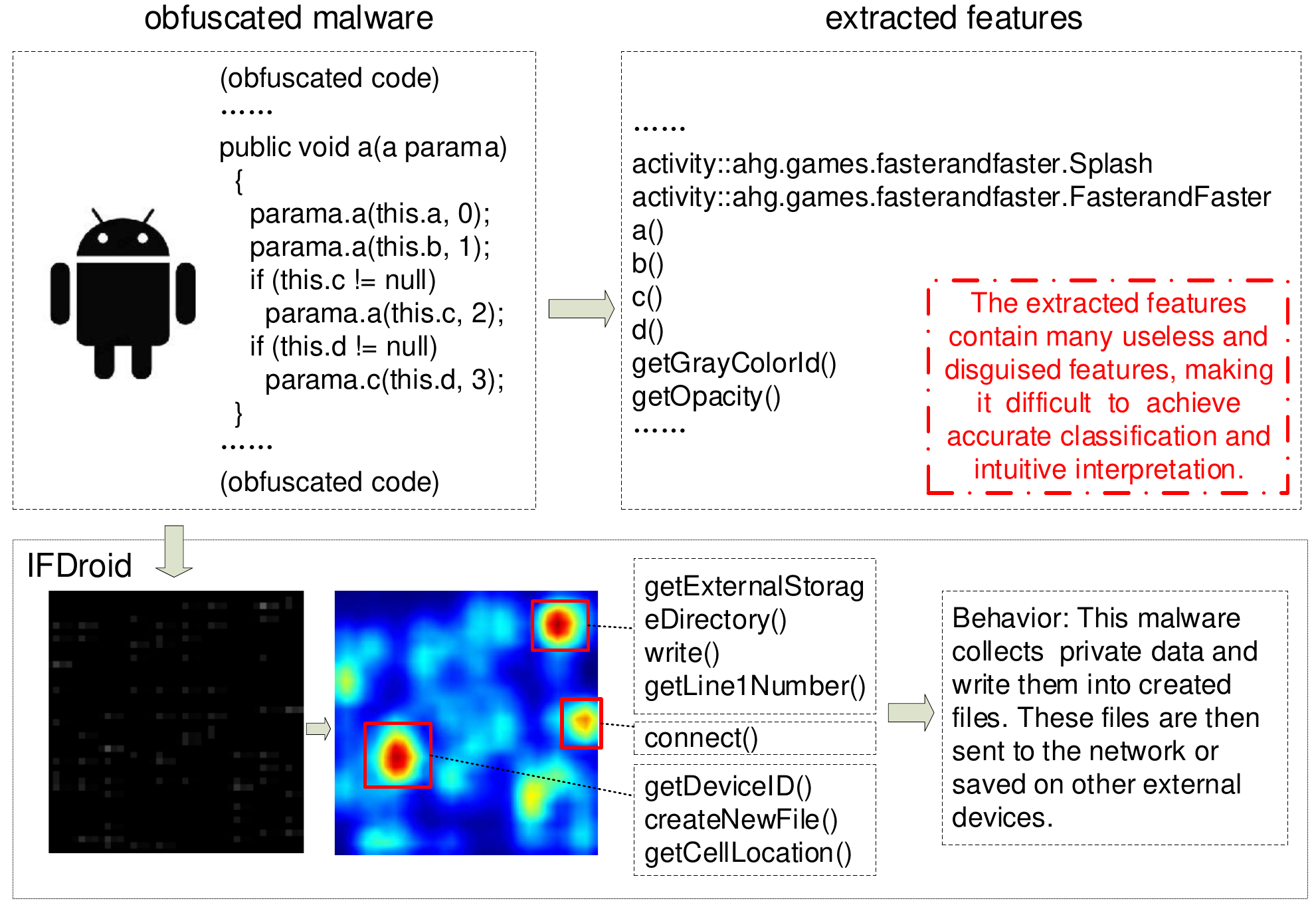}}
\caption{Motivation scenario of \emph{IFDroid}}
\label{fig:motivation}
\end{figure}

Assuming that there is a security analyst, his daily task is to classify newly detected malware into corresponding families to enrich their malware family dataset.
The richer the family dataset, the more accurately the unknown malware behavior can be predicted.
AV-TEST Institute \cite{AVTEST} reported that the average number of new malware detected per day is about 9,000. 
It will be a very time-consuming project if the analyst conducts in-depth manual analysis on these malware samples one by one.
Therefore, the analyst decides to extract the semantic information of the samples by program analysis, and then classify them into their corresponding families through semantic similarity matching (\eg graph matching).
But in fact, obfuscation technology has become more and more advanced, and it is used more and more frequently by attackers~\cite{2018obfuscate1, 2018obfuscate2}. 
In other words, these samples may be applied to different obfuscation techniques, resulting in the extracted features containing many useless and camouflage features.
At the same time, after classifying the samples into their families, the analyst wants to know which semantic features make these malware samples classified into corresponding families.
However, the classification method can only tell us which family the malware belongs to, and will not explain which features are used to determine that they are classified into this family.
The whole scenario is shown in Figure \ref{fig:motivation}.

To address the above challenges in the scenario, we first transform the program semantics of samples into images and then train a robust encoder by contrastive learning.
In classification phase, we use a visualization technique to obtain the corresponding heatmaps of generated images.
These heatmaps can help security analysts understand which features are more valuable in classifying them as corresponding families.
We implement \emph{IFDroid} to complete the whole analysis process automatically.

\section{System Architecture}
In this section, we introduce \emph{IFDroid}, a novel contrastive learning-based robust and interpretable Android malware classification system.

\begin{figure}[htbp]
\centerline{\includegraphics[width=0.49\textwidth]{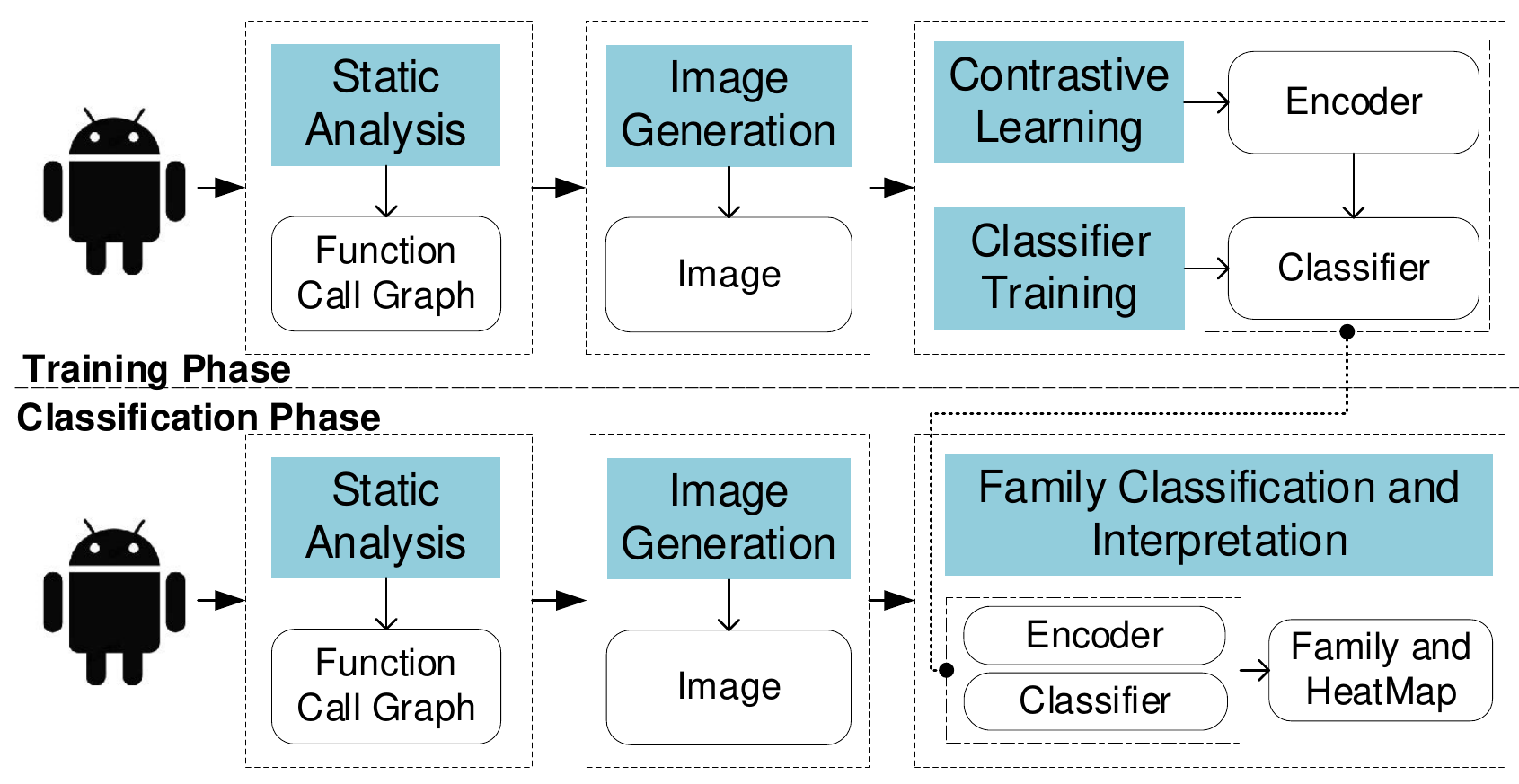}}
\caption{System overview of \emph{IFDroid}}
\label{fig:overview}
\end{figure}

\subsection{Overview}
\par As shown in Figure \ref{fig:overview}, \emph{IFDroid} consists of two main phases: \emph{Training Phase} and \emph{Classification Phase}.

The goal of \emph{Training Phase} is to train a robust encoder and an accurate classifier.
This phase consists of four steps as follows. 
\emph{1) Static Analysis}: This step aims to extract the function call graph of a malware sample based on static analysis where each node is an API call or a user-defined function.
\emph{2) Image Generation}: This step aims to transform the function call graph into an image based on centrality analysis.
\emph{3) Contrastive Learning}: This step aims to learn an encoder that can automatically extract robust features from an image.
\emph{4) Classifier Training}: This step aims to use vectors encoded by the learned encoder to train an accurate classifier.
    
The purpose of \emph{Classification Phase} is to classify unlabeled malware into their corresponding families.
This phase includes three steps: \emph{1) Static Analysis}, \emph{2) Image Generation}, and \emph{3) Family Classification and Interpretation}.
The first two steps are the same as in \emph{Training Phase}.
Given an image, it will be fed into a learned encoder in \emph{Training Phase} to obtain the vector representation.
The vector is then labeled as corresponding family by a trained classifier in \emph{Training Phase}. 
To interpret the classification result, we use a deep visualization technique to obtain the corresponding heatmap of the image to help the security analyst understand why it is classified as this family.

\subsection{Static Analysis}

Empirical studies~\cite{fan2018android, zhang2014droidsift} have demonstrated that graph representation is more robust than string-based features.
In this paper, we aim to achieve efficient malware analysis.
Therefore, we perform low-cost program analysis (\eg context- and flow-insensitive analysis) to distill the program semantics of a malware sample into a function call graph.
More specifically, we leverage a widely used Android reverse engineering tool namely Androguard~\cite{desnos2011androguard} to complete our static analysis.

To better describe the detailed steps in \emph{IFDroid}, we choose a real-world malware sample\footnote{485c85b5998bfceca88c6240e3bd5337} as our example. 
Figure \ref{fig:imagegenaration} shows the sample's function call graph where each node is an API call or a user-defined function. 
The number of nodes and edges are 117 and 170, respectively.

\subsection{Image Generation}

On the one hand, deep-learning-based image classification can process millions of images while maintaining high accuracy.
On the other hand, the output of image classification can be visualized to give a better intuition to users rather than giving a single decision. 
Because of these advantages, image-based methods have been widely used in malware analysis.
However, most of these approaches \cite{malware-image1, malware-image2, malware-image3, malware-image4} only use simple mapping algorithms to transform malware samples into images and then apply deep learning to analyze them.
Thus the semantics of the malware samples may be ignored.

To achieve efficient and semantic malware analysis, we use the technique (\ie centrality analysis) in our previous work~\cite{wu2019malscan} to transform the function call graph into an image.
The centrality concept was first proposed in social network analysis whose purpose is to dig out the most important persons in the network.
It can measure the importance of a node in a network and is very useful for network analysis.
In practice, there have been proposed many studies to use centralities in different areas such as biological network~\cite{jeong2001lethality}, co-authorship network~\cite{liu2005co}, transportation network~\cite{guimera2005worldwide}, criminal network~\cite{coles2001s}, etc

Different centralities analyze the importance of a node in a network by performing different network analyses, therefore, they have the potential to preserve different structural properties of a network.
In our paper, we select four widely used centrality measures (\ie \emph{Degree centrality}~\cite{freeman1978centrality}, \emph{Katz centrality}~\cite{katz1953new}, \emph{Closeness centrality}~\cite{freeman1978centrality}, and \emph{Harmonic centrality}~\cite{marchiori2000harmony}) to commence our image generation phase.
These four centralities can represent graph details from four different aspects. 
By this, we can achieve more complete preservation of a function call graph’s semantics.
Specifically, the definitions of these four centralities are as follows. 

\begin{itemize}

\item
\emph{Degree Centrality} \cite{freeman1978centrality} assigns an importance score based simply on the number of edges held by each node.
It is normalized by dividing by the maximum possible degree in a graph $N-1$, where $N$ denotes the number of nodes within the graph, $deg(i)$ is the degree of node $i$.  
\begin{equation}
C_{d}(i) = \frac{deg(i)}{N-1}
\end{equation}
\item 

\emph{Katz Centrality} \cite{katz1953new} computes the relative influence of a node within a graph by measuring the number of the immediate neighbors and also all other nodes in the graph that connect to the node under consideration through these immediate neighbors.
If $C_{k}(i)$ denotes Katz centrality of a node $i$, where the element at location $(i, j)$ of the adjacency matrix $A$ raised to the power $k$ (\ie $A^{k}$) reflects the total number of $k$ degree connections between nodes $i$ and $j$.
The $\alpha$ denotes an attenuation factor, then mathematically:
\begin{equation}
C_{k}(i)={\textstyle \sum_{k=1}^{\infty}}{\textstyle \sum_{j=1}^{\infty}} \alpha ^{k}(A^{k})_{ji}  
\end{equation}

\item \textbf{\emph{Closeness centrality}}~\cite{freeman1978centrality} indicates how close a node is to all other nodes in the network. 
It is calculated as the average of the shortest path length from the node to every other node in the graph. 
The smaller the average shortest distance of a node, the greater the closeness centrality of the node. 
In other words, the average shortest distance and the corresponding closeness centrality are negatively correlated. 
If $d(i,j)$ is the distance between nodes $i$ and $j$ and $N$ is the number of nodes in the graph, then mathematically:
\begin{equation}
C_c(i)=\frac{N-1}{\sum _{i\neq j} d(i,j)}
\end{equation}

\item \textbf{\emph{Harmonic centrality}}~\cite{marchiori2000harmony} reverses the sum and reciprocal operations in the definition of closeness centrality.
If $d(i,j)$ is the distance between nodes $i$ and $j$ and $N$ is the number of nodes in the graph, then mathematically:
\begin{equation}
C_h(i)=\frac{\sum \limits_{i\neq j}\frac{1}{d(i,j)}}{N-1}
\end{equation}

\end{itemize}

On the one hand, Android apps use API calls to access operating system functionality and system resources.
On the other hand, malware samples always invoke sensitive API calls to perform malicious tasks.
For example, \emph{getDeviceID} can get your phone's IMEI and \emph{getLine1Number} can obtain your phone number.
Therefore, we can leverage sensitive API calls to characterize the malicious behaviors of malware samples.
Specifically, we choose 426 sensitive API calls~\cite{Liangyi2020Experiences} as our concerned objectives which compose of three different API call sets. 
The first API call set is the top 260 API calls with the highest correlation with malware, the second API call set is 112 API calls that relate to restrictive permissions, and the third API call set is 70 API calls that are relevant to sensitive operations. 
Since the same API call may exist in different subsets, after calculating the union set of these three API call sets, the total number of API calls is not 442, but 426.

\begin{figure}[htbp]
\centerline{\includegraphics[width=0.485\textwidth]{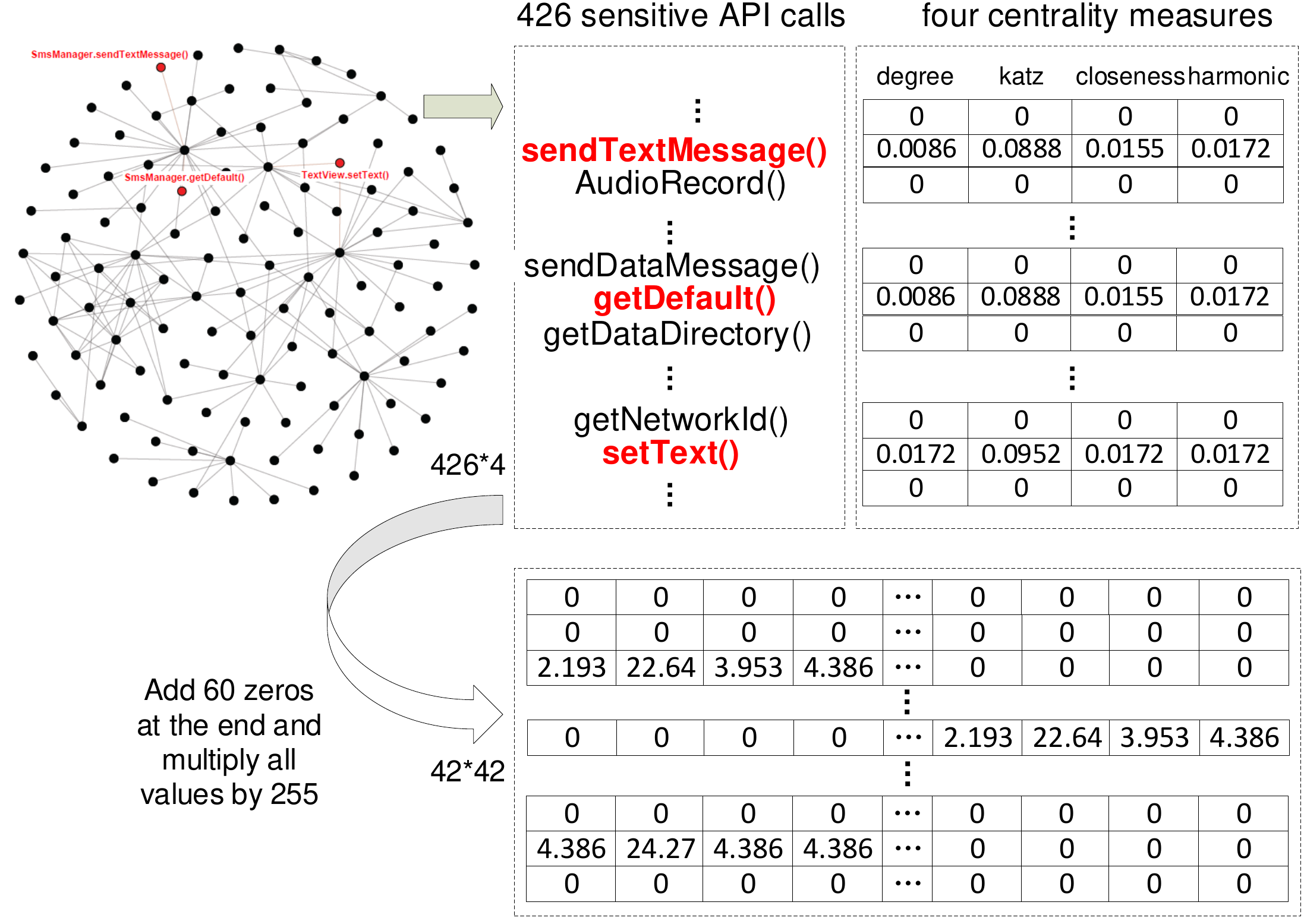}}
\caption{An example to illustrate the image generation step of \emph{IFDroid}}
\label{fig:imagegenaration}
\vspace{-1em}
\end{figure}

Given a function call graph, we first apply centrality analysis to obtain four centrality values of sensitive API calls.
If sensitive API calls do not appear in the function call graph, the four centrality values will all be zero. 
For example, the malware sample in Figure \ref{fig:imagegenaration} invokes a total of three sensitive API call (\ie \emph{sendTextMessage()}, \emph{getDefault()}, and \emph{setText()}). 
Then the four centrality values of these three sensitive API calls can be computed by centrality analysis.
Other 423 sensitive API calls do not appear in the call graph, therefore, their centrality values are all zero.
After centrality analysis, we can obtain a 426$*$4 vector representation.
If we directly transform it into an image, the generated image will be too narrow, making it difficult to distinguish which area it is when using heatmap for interpretation.
Moreover, it is not conducive to viewing.
Therefore, we crop the vector and turn it into a more square image.
Specifically, we add 60 zeros at the end and then reshape it as a 42$*$42 (\ie 426$*$4+60=42$*$42) vector.
At the same time, in order to be able to see more clearly, we multiply all the values in the vector by 255 to brighten the pixels in the image.
The range of centrality values is between zero and one, and the range of image pixels is between 0 and 255. 
After the two are multiplied, the range is also between 0 and 255.
Finally, we can obtain a 42$*$42 image.
Each pixel in the image represents a certain centrality value for a sensitive API call. 
Even if these pixels are separated, each pixel is still meaningful and can still represent the graph details of the sensitive API call.

\subsection{Contrastive Learning}

In our daily life, humans can recognize objects in the wild, even if we do not remember the exact appearance of the object. 
This happens because we have retained enough high-level features of the object to distinguish it from others and ignored pixel-level details.
For example, despite we have seen what a dollar bill looks like many times, we rarely draw a dollar bill exactly the same.
However, although we cannot draw a lifelike dollar bill, we can easily distinguish it~\cite{blog_contrastive}.
Therefore, researchers have asked a question: \emph{Can we build a representation learning algorithm that does not pay attention to pixel-level details and only encodes high-level features that are sufficient to distinguish different objects?}
To answer the question, contrastive learning is proposed.

The goal of contrastive learning is to maximize the agreement between original data and its positive data while minimize the agreement between original data and its negative data by using a contrastive loss in the vector space.
Note that $x$ is a sample, $x^{+}$ is a positive (\ie similar) sample of $x$, and $x^{-}$ is a negative (\ie dissimilar) sample of $x$.
Encoder $f$ can encode samples into vector representations.
$s$ is a function that computes the similarity between two vectors.
In self-supervised contrastive learning, the positive sample $x^{+}$ of an image $x$ is constructed by data augmentations such as image rotation and image cropping.
As for negative sample $x^{-}$, any other images can be selected as $x^{-}$.

\begin{figure}[htbp]
\centerline{\includegraphics[width=0.42\textwidth]{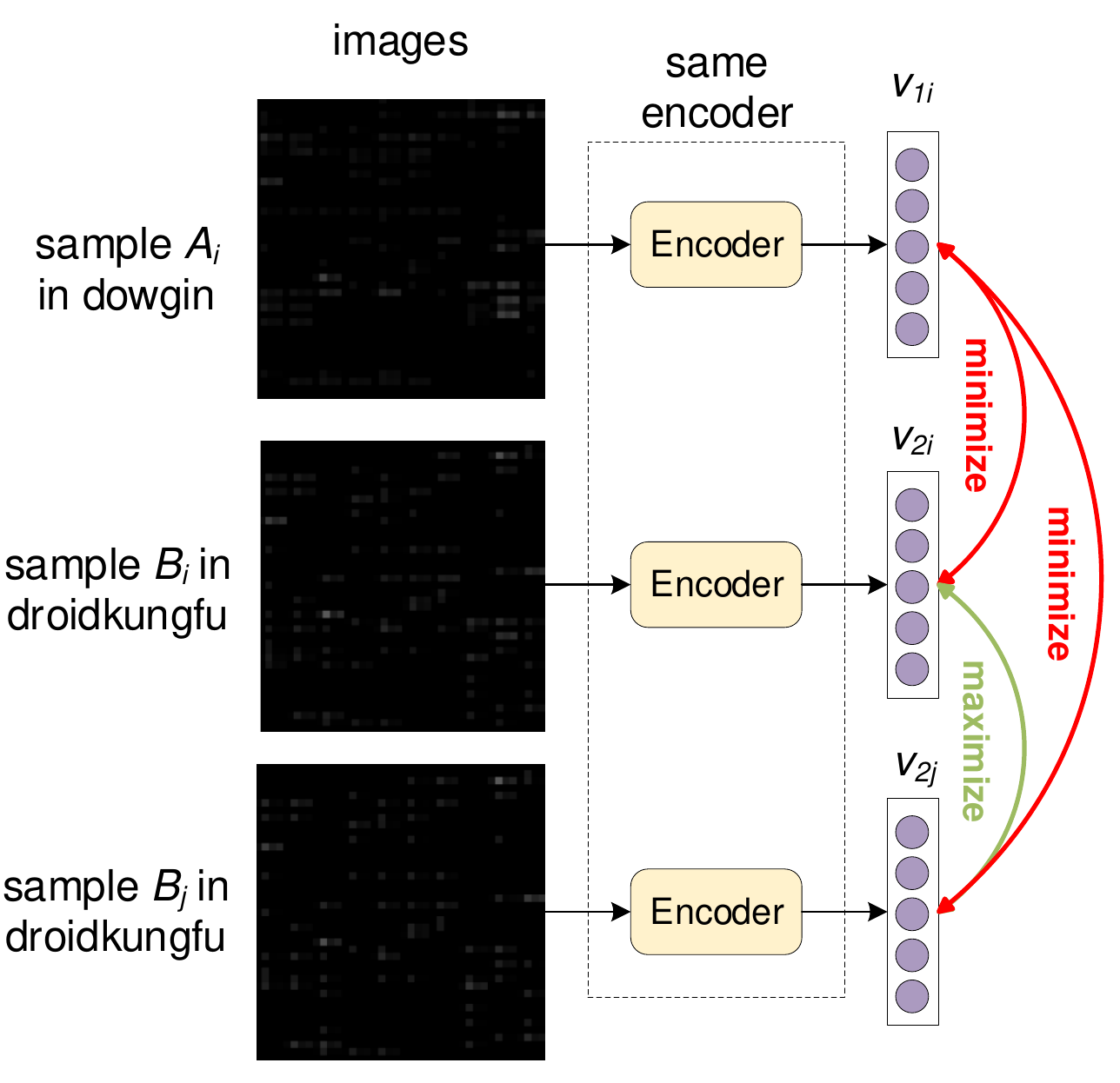}}
\caption{Supervised contrastive learning in \emph{IFDroid}, the goal is to maximize the similarity between samples in the same family and minimize the similarity between samples in different families.}
\label{fig:contrastive}
\end{figure}

Many studies have demonstrated the high effectiveness of self-supervised contrastive learning~\cite{2020DeCLUTR, 2020Cert, 2020InfoXLM}, but Khosla \emph{et al.} \cite{khosla2020supervised} found that the label information of samples can be used to improve the accuracy of contrastive learning.
Therefore, in this paper, we select supervised contrastive learning to train our encoder.
In other words, our positive samples $x^{+}$ are selected from other malware samples in the same family, rather than being data augmentations of itself, as done in self-supervised contrastive learning.
As shown in Figure \ref{fig:contrastive}, sample $B_{i}$ and $B_{j}$ are both from droidkungfu family while sample $A_{i}$ is from dowgin family.
Images of these three malware samples (\ie $A_{i}$, $B_{i}$, and $B_{j}$) are passed through an encoder to get their corresponding vector representations (\ie $v_{1i}$, $v_{2i}$, and $v_{2j}$).
Then the goal of our contrastive learning is to maximize the similarity between positive samples (\ie ($v_{2i}$, $v_{2j}$)) and minimize the similarity between negative samples (\ie ($v_{1i}$, $v_{2i}$) and ($v_{1i}$, $v_{2j}$)).

Specifically, supervised contrastive learning (SupCon) calculates contrastive loss in batches. 
Given an input batch of data, SupCon first applies data augmentation twice to obtain two copies of the batch and then combines these two augmented batches to form a multi-viewed batch. 
For each anchor sample $x_i$ with label $l_i$, the positive samples are all samples labeled $l_i$ in the same batch and the negative samples are the remaining samples whose labels are different from $l_i$. 
Within a multi-viewed batch of data, let $i \in I \equiv\{1,...,2N\}$ be the index of an arbitrary augmented sample, the loss takes the following form. 
\begin{equation}
    \mathcal{L}^{sup}_{out} = \sum_{i \in I} \frac{-1}{|P(i)|} \sum_{p \in P(i)}log\frac{exp(v_i \cdot v_p/\tau)}{\sum_{a \in A(i)}exp(v_i \cdot v_a/\tau)}	
    \label{equ:supcon}
\end{equation}
Here, $v_{i}$ represents the embedding of data $x_i$, $\tau \in \mathcal R^+$ is a scalar temperature parameter.
$A(i)\equiv I/\{i\}$ and $P(i)\equiv \{p\in A(i): l_p= l_i\}$ is the set of indices of all positives in the batch apart from $i$,  $|P(i)|$ is its cardinality.
The equation (\ref{equ:supcon}) uses multiple positive and negative samples per batch, and brings more information advantages to contrastive learning. 
It uses the positive normalization factor (\ie $\frac{1}{|P(i)|}$) to remove bias present in multiple positives samples and preserve the summation over negatives in the denominator to increase performance which has been shown in many studies \cite{he2020momentum, henaff2020data}.

\begin{table}[htbp]
\footnotesize
  \centering
  \caption{Parameters used in our contrastive learning}
    \begin{tabular}{|c|c|}
    \hline
    Parameters & Settings \\
    \hline
    loss function & SupCon loss in~\cite{khosla2020supervised} \\
    temperature & 0.07 \\
    optimizer & SGD \\
    momentum & 0.9 \\
    weight decay & 0.0001 \\
    learning rate & 0.05 \\
    batch size & 64 \\
    epoch & 100 \\
    \hline
    \end{tabular}%
  \label{tab:parameters1}%
\end{table}%

After experimenting with certain widely-used neural networks, we finally choose ResNet-18~\cite{he2016resnet18} as our image encoder since it can achieve a balance between accuracy and efficiency.
In computer vision tasks, the input of common datasets is 224$*$224 images with three channels. 
In our paper, the final image size is only 42$*$42, and there is only one channel. 
To make ResNet-18 suitable for our classification task, we modify the fully connected layer at the input of ResNet-18 so that the dimension of the input is the same as the size of our image. 
At the same time, in the whole process of ResNet-18, we reduce the depth to one-third of the original to adapt to our malware classification task.
Moreover, because contrastive learning requires the original sample, the corresponding positive samples, and negative samples when calculating loss, it is necessary to use batch normalization to normalize the output of different samples to ensure that the magnitude of the output of all samples is consistent. 
In other words, we use the normalized output to train our classifier.
Table \ref{tab:parameters1} shows the details of parameters used in our contrastive learning.
The loss function is the same as in \emph{SupCon}~\cite{khosla2020supervised} namely \emph{Supervised Contrastive Loss}.
The whole procedure is trained using \emph{Stochastic Gradient Descent} (SGD) with 0.9 momentum and 0.0001 weight decay.
The output in this step is a learned encoder, that is, a learned ResNet-18.
It can convert an image into a vector whose dimension is 512.

\subsection{Classifier Training}

In this step, we first use our learned encoder (\ie ResNet-18) to encode images into corresponding vectors, and then train a classifier (\ie a one-layer fully connected layer) by using these vectors and their labels.
After training 100 epochs, the classifier will be selected as our final classifier.
Parameters used in classifier training and contrastive learning are different only in loss function (\ie \emph{Cross Entropy in classifier training}), and the others are the same.

\subsection{Family Classification and Interpretation}

\begin{figure}[htbp]
\centerline{\includegraphics[width=0.4\textwidth]{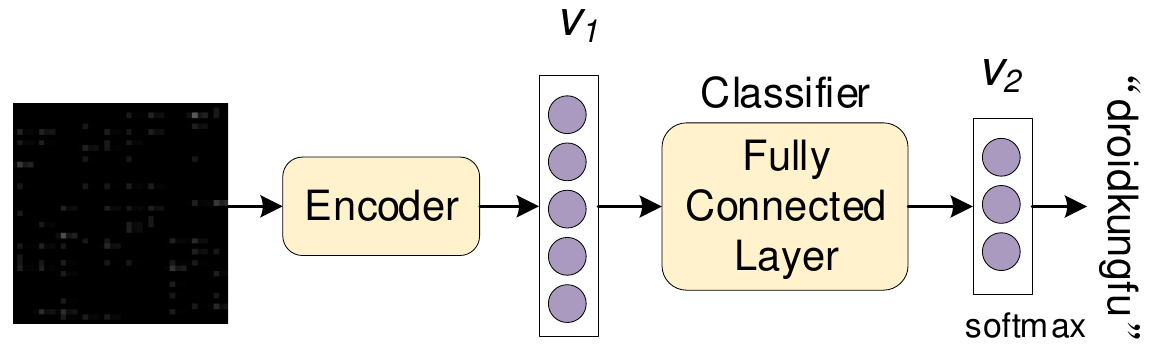}}
\caption{Familial classification of \emph{IFDroid}}
\label{fig:cls}
\end{figure}

After training phase, we can obtain a learned encoder and a trained classifier.
They will be used to classify newly unlabeled malware samples.
Specifically, given a new malware sample, we first perform static analysis to extract the function call graph.
Then the graph is transformed into an image by centrality analysis.
As shown in Figure \ref{fig:cls}, given an image, the encoder can encode it as a vector representation.
Finally, the classifier takes the input of the vector and predicts the corresponding family.

\begin{figure}[htbp]
\centerline{\includegraphics[width=0.46\textwidth]{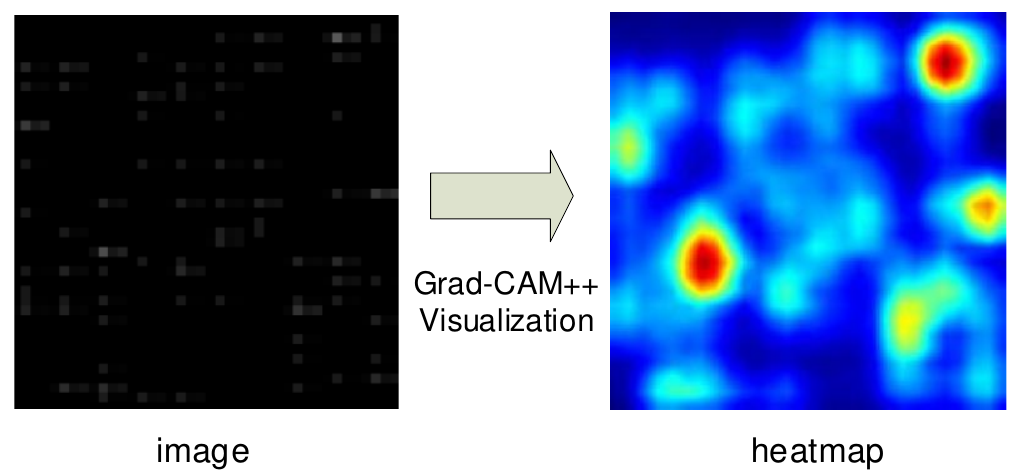}}
\caption{The heatmap generated by Grad-CAM++ visualization technique to illustrate the classification result
}
\label{fig:visualization1}
\end{figure}

Since we transform a function call graph into an image and leverage a CNN model to classify malware, we can apply visualization techniques to interpret the classification result. 
In practice, there have been strong efforts focusing on creating meaningful heatmaps that highlight the importance of individual pixel regions in an input image to its classification using a CNN. 
After trying several visualization techniques (\eg CAM \cite{zhou2016learningcam}, Grad-CAM \cite{selvaraju2017gradcam}, and Grad-CAM++ \cite{chattopadhay2018gradcam}), we find that Grad-CAM++ can achieve better interpretability. 
Particularly, it introduces pixel-wise weighting of the gradients of the output with respect to a particular spatial position in the final convolutional feature map of the CNN. 
In this way, it can provide a measure of the importance of each pixel in a feature map towards the overall decision of the CNN. 
Moreover, the visual explanations of Grad-CAM++ for any CNN-based network can be generated without changing the architecture or retraining. 
Therefore, we select it to interpret our classification result. 

Figure \ref{fig:visualization1} shows the visualization of a real-world malware sample.
According to the intensity of the color in the heatmap, we can know which features are more effective in classifying this malware as this family.
For example, a malware sample is classified as droidkungfu family.
After applying Grad-CAM++ visualization technique on the image, we can obtain the corresponding heatmap as shown in Figure \ref{fig:visualization1}.
The redder area in the heatmap indicates that this area contains more effective features.
We can find the corresponding areas in the original image in reverse. 
Sensitive API calls contained in these areas have higher weights in identifying the malware as the family.

\section{Experimental Evaluation}
\par In this section, we aim to answer the following research questions:

\begin{itemize}
  \item \emph{RQ1: What is the effectiveness of IFDroid on classifying Android malware without and with obfuscations?}
  \item \emph{RQ2: Can IFDroid interpret the familial classification results?}
  \item \emph{RQ3: What is the runtime overhead of IFDroid on classifying Android malware?}
\end{itemize}

\subsection{Datasets and Metrics}

\begin{table}
  \footnotesize
  \centering
\caption{Descriptions of metrics used in our experiments}
  \begin{tabular}{|m{2.3cm}<{\centering}|m{0.6cm}<{\centering}|m{4.5cm}<{\centering}|}
  \hline
    \textbf{Metrics}&\textbf{Abbr}&\textbf{Definition}\\
    \hline
    True Positive & \textbf{TP} & \#malware in family \emph{f} are correctly classified into family \emph{f} \\
    True Negative & \textbf{TN} & \#malware not in family \emph{f} are correctly not classified into family \emph{f} \\
    False Positive & \textbf{FP} &  \#malware not in family \emph{f} are incorrectly classified into family \emph{f} \\
    False Negative & \textbf{FN} &  \#malware in family \emph{f} are incorrectly not classified into family \emph{f} \\
    True Positive Rate & \textbf{TPR} & TP/(TP+FN)\\
    False Negative Rate & \textbf{FNR} & FN/(TP+FN)\\
    True Negative Rate & \textbf{TNR}& TN/(TN+FP)\\
    False Positive Rate & \textbf{FPR} & FP/(TN+FP)\\
    Precision & \textbf{P} & TP/(TP+FP)\\
    Recall & \textbf{R} & TP/(TP+FN)\\
    F-measure & \textbf{F1} & 2*P*R/(P+R)\\
    \hline
    Classification Accuracy & \textbf{CA} & percentage of malware which are correctly classified into their families\\
    \hline
  \end{tabular}

  \label{tab:metrics}
\end{table}

We first select one widely used ground truth dataset (\ie dataset-I~\cite{zhou2012dissecting}) as our experimental dataset to evaluate \emph{IFDroid}. 
It is provided by Genome project~\cite{zhou2012dissecting} which consists of 1,247 malware samples.
In fact, these malware samples were developed in 2011-2012, and are too old.
To achieve more comprehensive evaluations, we construct another new larger dataset.
Specifically, we choose the malware samples in AndroZoo \cite{allix2016androzoo} as our targets since most of them have been scanned by antivirus products in VirusTotal \cite{virustotal}.
After collecting all scanning reports, these malware samples can be labeled into their corresponding families by using a technique namely \emph{Euphony} \cite{hurier2017euphony}.
To construct our new dataset-II, we randomly download 8,000 samples from 15 largest families and the number of samples in our new dataset is 120,000. 
Note that AndroZoo is one of the largest Android app collections which contains more than 10 million Android apps.
\emph{Euphony} is a tool to label the family of a malware sample by analyzing the VirusTotal reports and has been used by many other researchers~\cite{zhu2020measuring, zhang2019familial}.

To evaluate \emph{IFDroid}, we conduct experiments by performing ten-fold cross-validations. 
In other words, we first divide our dataset into ten subsets, then nine of them are selected as training sets and the last subset is used to test.
We repeat this ten times and report the average classification results.
Furthermore, to measure the effectiveness of \emph{IFDroid}, we leverage certain widely used metrics (Table \ref{tab:metrics}) to present the classification results.
We run our experiments on a desktop equipped with a 32-core 2.30GHz CPU, a GTX 1080 GPU, and 128GB of RAM.

\begin{table}[htbp]
\footnotesize
  \centering
  \caption{Classification accuracy of \emph{IFDroid} and six state-of-the-art comparative systems on dataset-I~\cite{zhou2012dissecting}}
    \begin{tabular}{|c|c|}
    \hline
    Baseline Approach & Classification Accuracy \\
    \hline
    Dendroid~\cite{suarez2014dendroid} & 0.942 \\
    Apposcopy~\cite{feng2014apposcopy} & 0.900 \\
    DroidSIFT~\cite{zhang2014droidsift} & 0.930 \\
    MudFlow~\cite{avdiienko2015mudflow} & 0.881 \\
    DroidLegacy~\cite{deshotels2014droidlegacy} & 0.929 \\
    Astroid~\cite{feng2016astroid} & 0.938 \\
    \textbf{IFDroid} & \textbf{0.984} \\
    \hline
    \end{tabular}%
  \label{tab:dataset1_result}%
\end{table}%

\subsection{RQ1: Effectiveness}

\subsubsection{\textbf{Effectiveness on Classifying General Malware}}

First, we conduct evaluations to check the ability of \emph{IFDroid} on classifying general Android malware.
We first use dataset-I to present the comparative results of \emph{IFDroid} and six state-of-the-art related systems.
These systems include: 1)  \emph{Dendroid}~\cite{suarez2014dendroid} applies text mining techniques to analyze the code structures of Android malware and classify them into corresponding families;
2) \emph{Apposcopy}~\cite{feng2014apposcopy} performs program analysis to extract both data-flow and control-flow information of malware samples to classify them;
3) \emph{DroidSIFT}~\cite{zhang2014droidsift} pays attention to constructing API dependency graph by analyzing the program semantics to classify Android malware;
4) \emph{MudFlow}~\cite{avdiienko2015mudflow} extracts the source-and-sink pairs of malware samples and regards them as features to classify malware;
5) \emph{DroidLegacy}~\cite{deshotels2014droidlegacy} conducts app partition to divide the malware sample into sub-modules and labels the corresponding family by analyzing the malicious sub-module;
and 6) \emph{Astroid}~\cite{feng2016astroid} synthesizes a maximally suspicious common subgraph of each malware family as a signature to classify malware.

Because most of these systems are not publicly available and the dataset used to produce evaluation results are all the same (\ie dataset-I~\cite{zhou2012dissecting}), we directly adopt the classification accuracy of these six systems in their papers~\cite{suarez2014dendroid, feng2014apposcopy, zhang2014droidsift, avdiienko2015mudflow, deshotels2014droidlegacy, feng2016astroid}.
Through results in Table \ref{tab:dataset1_result}, we see that \emph{IFDroid} is superior to other comparative systems.
This happens because \emph{IFDroid} not only considers the program semantics of malware samples but also extracts effective features by a learned robust encoder.

\begin{table}[htbp]
  \centering
  \footnotesize
 \caption{Familial classification results (F1) of \emph{FalDroid} (\emph{Fal} for short), \emph{AOM}, \emph{MVIIDroid} (\emph{MVI} for short), \emph{CDFG}, and \emph{IFDroid} on dataset-II.
 \emph{wo} and \emph{wi} denote \emph{IFDroid} without and with contrastive learning, respectively.}
    \begin{tabular}{|m{2cm}<{\centering}|m{0.62cm}<{\centering}m{0.62cm}<{\centering}m{0.62cm}<{\centering}m{0.65cm}<{\centering}|m{0.62cm}<{\centering}m{0.62cm}<{\centering}|}
    \hline
    Family & \emph{Fal} & \emph{AOM}  & \emph{MVI} & \emph{CDFG}  & \emph{wo}  & \emph{wi} \\
    \hline
    adwo  & 90.5  & 83.2  & 85.3  & 91.6  & 91.7  & \textbf{93.5} \\
    airpush & 76.3  & 64.9  & 70.1  & 82.3  & 78.5  & \textbf{92.5} \\
    dowgin & 95.2  & 84.5  & 89.1  & 96.1  & 97.1  & \textbf{97.5} \\
    droidkungfu & 94.7  & 80.4  & 89.1  & 96.2  & 98.7  & \textbf{99.5} \\
    feiwo & 94.8  & 91.3  & 93    & \textbf{95.2} & 93.1  & 94.4 \\
    gingermaster & 92.4  & 83.9  & 90.8  & 93.5  & 95.1  & \textbf{97.3} \\
    kuguo & 92.5  & 89.3  & 91.1  & 93.6  & 93.1  & \textbf{94.9} \\
    leadbolt & 95.2  & 93.3  & 94.1  & 96.7  & 98.8  & \textbf{99.4} \\
    plankton & 99.1  & 91.4  & 95.2  & 98.6  & 98.2  & \textbf{99.1} \\
    startapp & 90.4  & 87.2  & 90.3  & 95.7  & 99.2  & \textbf{100} \\
    umeng & 90.3  & 82.5  & 87.1  & 92.1  & 90.3  & \textbf{93.2} \\
    utchi & 100   & 100   & 100   & 100   & 100   & \textbf{100} \\
    waps  & 93.3  & 91.3  & 93.9  & 96.2  & 95.8  & \textbf{96.5} \\
    wooboo & 92.4  & 88.3  & 90.2  & 97.3  & 97.1  & \textbf{98.2} \\
    youmi & 78.4  & 73.5  & 77.1  & 83.6  & 85.9  & \textbf{88.7} \\
    \hline
    \end{tabular}%
  \label{tab:dataset2_result2}%
\end{table}%

To achieve more comprehensive evaluations, we use our new dataset-II to compare \emph{IFDroid} with four recent state-of-the-art Android malware familial classification methods.
They include: 
1) \emph{FalDroid}~\cite{fan2018android} analyzes frequent subgraphs to represent the common behaviors of each malware family and uses them to perform familial classification;
2) \emph{AOM}~\cite{blanc2019identifyingAOM} uses Android-oriented metrics to identify Android malware families;
3) \emph{MVIIDroid}~\cite{wu2020mviidroidMVI} uses a multiple view information integration approach for Android malware detection and family identification;
and 4) \emph{CDFG}~\cite{zhiwu2019androidCDFG} combines control flow graph with data flow graph to accomplish Android malware family classification.
To further examine the contribution of contrastive learning to general malware classification, we implement another system that does not use contrastive learning to learn an encoder first but directly trains the encoder and classifier together in training phase.
We name this system \emph{IFDroid} (\emph{wo}) because it differs from \emph{IFDroid} in that it is trained \emph{w}ith\emph{o}ut using contrastive learning.

\begin{table}[htbp]
\footnotesize
  \centering
  \caption{Descriptions of 12 obfuscators used in our experiments}
    \begin{tabular}{|c|m{4.8cm}<{\centering}|}
    \hline
    Obfuscators  & Descriptions \\
    \hline
    ClassRename    & Change the package name and rename classes \\
    FieldRename      & Rename fields \\
    MethodRename      & Rename methods \\
    ConstStringEncryption     & Encrypt constant strings in code \\
    AssetEncryption     & Encrypt asset files \\
    LibEncryption      & Encrypt native libs \\
    ResStringEncryption      & Encrypt strings in resources \\
    ArithmeticBranch     & Insert junk code that is composed by arithmetic computations and a branch instruction \\
    CallIndirection    & Modify the control-flow graph without changing the code semantics \\
    Goto     & Modify the control-flow graph by adding two new nodes \\
    Nop     & Insert random nop instructions within every method implementation \\
    Reorder     & Change the order of basic blocks of the control-flow graph \\
    \hline
    \end{tabular}%
  \label{tab:obfuscators}%
\end{table}%

\begin{table*}[htbp]
\footnotesize
  \centering
  \caption{F1 values of \emph{FalDroid} (\emph{Fal} for short), \emph{AOM}, \emph{MVIIDroid} (\emph{MVI} for short), \emph{CDFG}, and \emph{IFDroid} on classifying obfuscated malware. \emph{wo} and \emph{wi} denote \emph{IFDroid} without and with contrastive learning, respectively.}
    \begin{tabular}{|cc|c|c|c|c|c|c|c|}
    \hline
    \multicolumn{2}{|c|}{Obfuscators} & \#Samples & \textit{Fal} & \textit{AOM} & \textit{MVI} & \textit{CDFG} & \textit{wo} & \textit{wi} \\
    \hline
    \multicolumn{1}{|c|}{\multirow{3}[2]{*}{Rename}} & \multicolumn{1}{c|}{ClassRename} & 5,049 & 100.0 & 99.1  & 99.7  & 100.0 & 100.0 & \textbf{100.0} \\
    \multicolumn{1}{|c|}{} & \multicolumn{1}{c|}{FieldRename} & 5,297 & 100.0 & 98.4  & 99.8  & 100.0 & 100.0 & \textbf{100.0} \\
    \multicolumn{1}{|c|}{} & \multicolumn{1}{c|}{MethodRename} & 5,271 & 100.0 & 98.6  & 99.6  & 100.0 & 100.0 & \textbf{100.0} \\
    \hline
    \multicolumn{1}{|c|}{\multirow{4}[2]{*}{Encryption}} & \multicolumn{1}{c|}{AssetEncryption} & 5,477 & 93.8  & 82.1  & 91.2  & 96.5  & 98.1  & \textbf{100.0} \\
    \multicolumn{1}{|c|}{} & \multicolumn{1}{c|}{ConstStringEncryption} & 5,443 & 84.7  & 71.4  & 79.3  & 88.3  & 90.2  & \textbf{96.9} \\
    \multicolumn{1}{|c|}{} & \multicolumn{1}{c|}{LibEncryption} & 5,391 & 94.5  & 87.1  & 91.9  & 96.8  & 96.9  & \textbf{100.0} \\
    \multicolumn{1}{|c|}{} & \multicolumn{1}{c|}{ResStringEncryption} & 5,366 & 95.1  & 91.4  & 93.6  & 99.1  & 98.0  & \textbf{100.0} \\
    \hline
    \multicolumn{1}{|c|}{\multirow{5}[2]{*}{Code}} & \multicolumn{1}{c|}{ArithmeticBranch} & 5,649 & 97.1  & 82.4  & 87.1  & 97.5  & 97.3  & \textbf{100.0} \\
    \multicolumn{1}{|c|}{} & \multicolumn{1}{c|}{CallIndirection} & 5,571 & 73.2  & 62.4  & 69.3  & 77.1  & 77.3  & \textbf{91.1} \\
    \multicolumn{1}{|c|}{} & \multicolumn{1}{c|}{Goto} & 5,546 & 92.3  & 82.1  & 90.4  & 92.6  & 94.8  & \textbf{100.0} \\
    \multicolumn{1}{|c|}{} & \multicolumn{1}{c|}{Nop} & 5,529 & 95.8  & 90.2  & 90.8  & 95.6  & 97.7  & \textbf{100.0} \\
    \multicolumn{1}{|c|}{} & \multicolumn{1}{c|}{Recorder} & 5,443 & 96.0  & 86.3  & 92.3  & 96.1  & 95.5  & \textbf{100.0} \\
    \hline
    \multicolumn{2}{|c|}{Apply 12 obfuscators} & 4,389 & 69.2  & 53.5  & 65.3  & 72.6  & 71.8  & \textbf{90.4} \\
    \hline
    \multicolumn{2}{|c|}{ALL } & 69,421 & 91.9  & 83.7  & 88.7  & 93.5  & 93.9  & \textbf{98.4} \\
    \hline
    \end{tabular}%
  \label{tab:obfuscation}%
\end{table*}%

The detailed classification results of 15 families are shown in Table \ref{tab:dataset2_result2}.
Through the results, we observe that \emph{IFDroid} performs better than \emph{FalDroid}, \emph{AOM}, \emph{MVIIDroid}, and \emph{CDFG} on classifying most malware families.
For example, the F1 values of \emph{FalDroid}, \emph{AOM}, \emph{MVIIDroid}, and \emph{CDFG} on classifying ``startapp" family are 90.4\%, 87.2\%, 90.3\%, and 95.7\%, respectively.
However, for \emph{IFDroid}, it has the ability to classify them into ``startapp" family without any inaccuracies.
Additionally, when we use contrastive learning to learn our encoder first, the classification performance is always better than without contrastive learning.
For example, when classifying malware samples with ``airpush'' family, the F1 value of \emph{IFDroid} (\emph{wo}) is only 78.5\% while \emph{IFDroid} with contrastive learning can maintain 92.5\% F1.
Such results suggest that the use of contrastive learning can indeed enhance the ability of \emph{IFDroid} on classifying malware.

\subsubsection{\textbf{Effectiveness on Classifying Obfuscated Malware}}

Next, we evaluate the effectiveness of \emph{IFDroid} on classifying obfuscated Android malware.
For this purpose, we use an automatic Android apps obfuscation tool (\emph{Obfuscapk}~\cite{aonzo2020obfuscapk}) that provides certain obfuscators including typical obfuscations (\eg class rename and method rename) and some advanced code obfuscations (\eg call indirection and goto).
Specifically, we first learn an encoder and train a classifier by using samples in dataset-II.
After completing the training phase, we randomly select 400 malware samples from each family, and the final number of selected samples is 400$*$15=6,000.

We select a total of 12 different obfuscators provided by \emph{Obfuscapk}, including three rename obfuscators, four encryption obfuscators, and five advanced code obfuscators.
Descriptions of these obfuscators are presented in Table \ref{tab:obfuscators}.
In practice, \emph{Obfuscapk} can not obfuscate some of our malware samples due to certain errors.
But fortunately, the failed samples only occupy a small part.
To further evaluate the effectiveness of \emph{IFDroid}, we apply 12 obfuscators together to generate more complex obfuscated malware.
Finally, we obtain 69,421 obfuscated samples in total.
We also conduct comparative evaluations with \emph{FalDroid}, \emph{AOM}, \emph{MVIIDroid}, and \emph{CDFG} on classifying these obfuscated malware.
The comparative results are shown in Table \ref{tab:obfuscation}, which includes the F1 values of \emph{IFDroid} and comparative methods (\ie \emph{FalDroid}, \emph{AOM}, \emph{MVIIDroid}, and \emph{CDFG}).

Since the typical rename obfuscations (\ie class rename, method rename, and field rename) do not change the call relationships between functions in an app.
Both \emph{FalDroid}, \emph{CDFG}, and \emph{IFDroid} can correctly classify all obfuscated apps into corresponding families.
Furthermore, no matter what kind of obfuscation it is for, \emph{IFDroid} can perform better when using contrastive learning.
However, the F1 values of \emph{IFDroid} without contrastive learning is only 77.3\% when we classify apps that are obfuscated by \emph{CallIndirection}.
After our in-depth analysis, we find that the number of nodes and edges in a function call graph change a lot after applying \emph{CallIndirection}.
For example, a sample originally has 8,135 nodes and 19,725 edges. 
After it is obfuscated by \emph{CallIndirection}, the number of nodes becomes 35,396, and the number of edges increases to 58,407. 
This huge change makes \emph{IFDroid} make mistakes in classification.
However, when we adopt contrastive learning, the true positive rate of \emph{IFDroid} can be increased from 77.3\% to 91.1\%. 
This result also shows that the encoder learned by contrastive learning can extract more robust features to classify malware.

Moreover, when classifying samples that are obfuscated by 12 obfuscators together, \emph{IFDroid} can also achieve an F1 of 90.4\%.
However, the F1 values of \emph{FalDroid}, \emph{AOM}, \emph{MVIIDroid}, and \emph{CDFG} are only 67.9\%, 50.1\%, 61.6\%, and 69.3\%, respectively.
On average, \emph{IFDroid} maintains an F1 of 98.4\% on classifying all generated obfuscated samples.
Such result suggests that \emph{IFDroid} is more robust than \emph{FalDroid}, \emph{AOM}, \emph{MVIIDroid}, and \emph{CDFG}.

\emph{\textbf{Summary:} 
IFDroid achieves higher accuracy than Dendroid, Apposcopy, DroidSIFT, MudFlow, DroidLegacy, Astriod, FalDroid, AOM, MVIIDroid, and CDFG on Android malware classification. 
Using contrastive learning can not only improve the accuracy but also enhance the robustness of IFDroid.}

\begin{figure}[htbp]
\centerline{\includegraphics[width=0.4\textwidth]{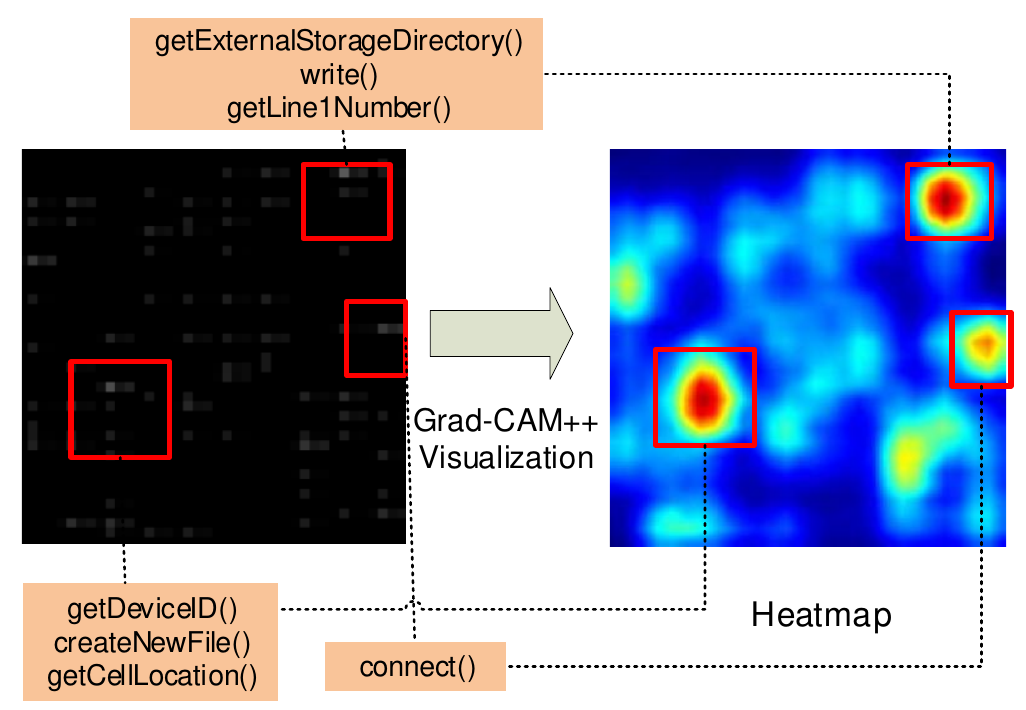}}
\caption{A real-world malware sample's visualization of classification result}
\label{fig:visualization2}
\end{figure}

\begin{figure*}[htbp]
\centerline{\includegraphics[width=0.99\textwidth]{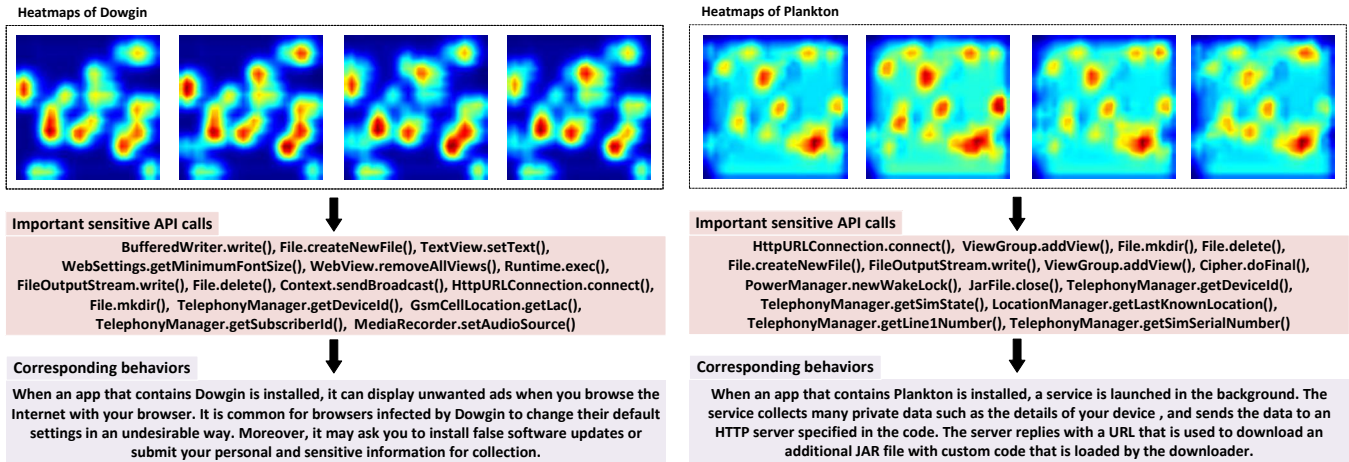}}
\caption{The most important sensitive API calls of two families obtained by analyzing the heatmaps}
\label{fig:behaviors}
\vspace{-1em}
\end{figure*}

\subsection{RQ2: Interpretability}

As aforementioned, since \emph{Gradient-weighted Class Activation Mapping++} (Grad-CAM++)~\cite{selvaraju2017gradcam, chattopadhay2018gradcam} performs better and can generate visual explanations for any CNN-based network without changing the architecture or retraining, we use it as our visualization technique to interpret our classification results.
These interpretations can help security analysts understand why a malware sample is classified as this family. 
Specifically, we use Grad-CAM++ to obtain the corresponding heatmaps of malware samples.
The red area in the heatmap indicates that features in this area play more important roles in classifying as this family.
In other words, sensitive API calls contained in these areas have higher weights in identifying the malware as the family.

Figure \ref{fig:visualization2} shows the visualization of a real-world malware sample.
This sample is correctly classified into droidkungfu family by \emph{IFDroid}.
After applying Grad-CAM++ visualization technique on the generated image, we can obtain the corresponding heatmap as shown in Figure \ref{fig:visualization2}.
The redder the color in a heatmap, the more valuable the features in the area.
Therefore, we pay more attention to these red areas.
After completing the one-to-one correspondence of three red areas from the original image, we can collect seven sensitive API calls (\ie \emph{getExternalStorageDirectory()}, \emph{write()}, \emph{getLine1Number()}, \emph{getDeviceID()}, \emph{createNewFile()}, \emph{getCellLocation()}, and \emph{connect()}) from these areas.
Through the result, we can see that this malware collects users' private data and write them into created files.
These files are then sent to the network or saved on other external devices.
Because of this behavior, it is classified into ``droidkungfu'' family.

\begin{figure}[htbp]
\centerline{\includegraphics[width=0.48\textwidth]{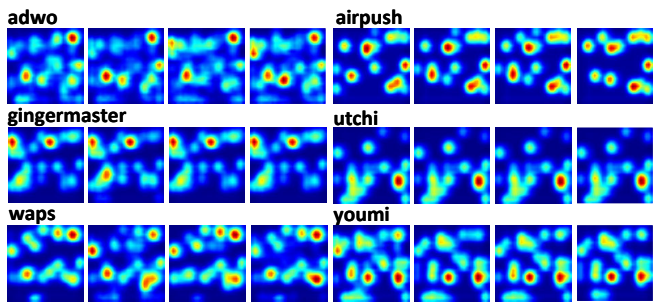}}
\caption{Visualization of six malware familial classification results, four malware samples are displayed for each family.}
\label{fig:visualization6family}
\end{figure}

To further analyze the interpretability of \emph{IFDroid}, we collect the most important sensitive API calls of each family by analyzing their heatmaps.
Due to space limitations, we only show the details of two families (\ie ``dowgin'' and ``plankton'') in Figure \ref{fig:behaviors}.
From the results in Figure \ref{fig:behaviors}, we see that the ``dowgin'' family behaves by displaying unwanted or malicious advertisements and changing web settings in undesirable ways.
For malware of the "plankton" family, once a user installs it, it can collect a lot of sensitive data such as device ID and send them to a remote server.
Meanwhile, the server replies with a URL that allows the infected phone to download and install a JAR file containing a dynamic payload.
Such payload can have a huge impact on users.

\begin{figure}[htbp]
\centerline{\includegraphics[width=0.45\textwidth]{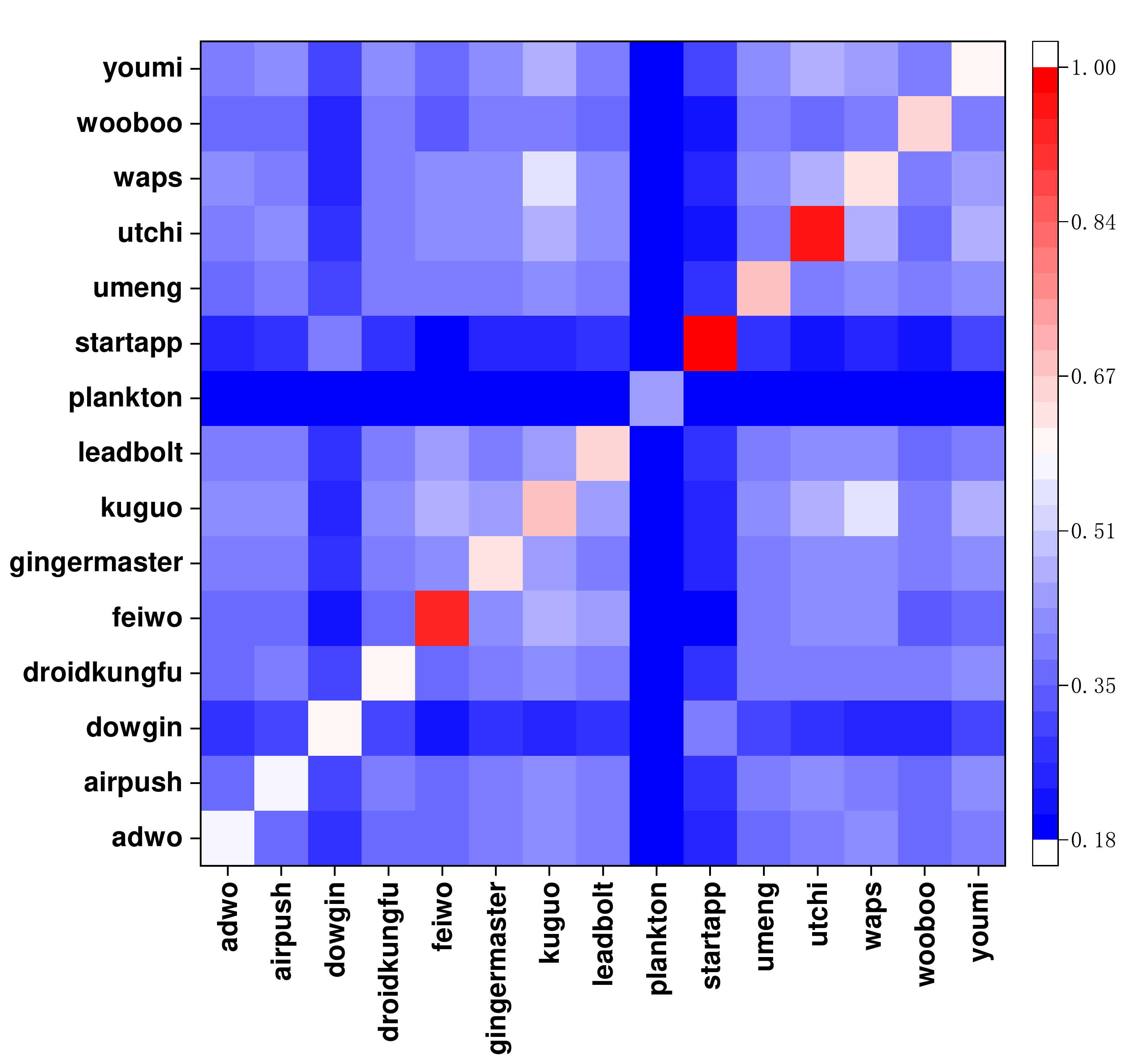}}
\caption{The average similarity between heatmaps in 15 families}
\label{fig:heatmap-sim}
\end{figure}

Finally, we conduct a survey to explore the effectiveness of the interpretation of family classification by heatmaps.
Due to the limited space, we randomly present six families' heatmaps in Figure \ref{fig:visualization6family}, and four malware samples are displayed for each family.
Given heatmaps of all malware samples, we compute the similarity of two heatmaps by using \emph{Structural SIMilarity} (SSIM)~\cite{wang2004ssim} technique one by one.
SSIM is widely used to measure the similarity of two images.

After analyzing all heatmaps, the average similarity between different families is calculated and presented in Figure \ref{fig:heatmap-sim}.
From Figure \ref{fig:visualization6family} and Figure \ref{fig:heatmap-sim}, we observe several phenomenons.
The first phenomenon is that the heatmaps of most malware in the same family are similar.
It is reasonable because malware samples of certain families are often just repackaged applications with slight modifications~\cite{arp2014drebin}.
In other words, most malware samples in the same family are polymorphic variants of other malware samples in this family.
Therefore, malware samples in the same family always perform similar malicious actions, resulting in similar heatmaps generated by visualization.
The second phenomenon is that the heatmaps of malware in different families are basically different.
This result is in line with expectations since malware samples in different families exhibit different malicious behaviors.
Because of this, they are classified into different malware families.

\emph{\textbf{Summary:} 
IFDroid can interpret the familial classification results by using visualization techniques. 
We can even distinguish the malware families directly through the heatmaps since the heatmaps of most malware in the same family are similar and heatmaps of malware in different families are basically different.}

\subsection{RQ3: Efficiency}

\begin{figure}
\centering
\subfigure{
\begin{minipage}[t]{0.22\textwidth}
\centering
\includegraphics[width=\textwidth]{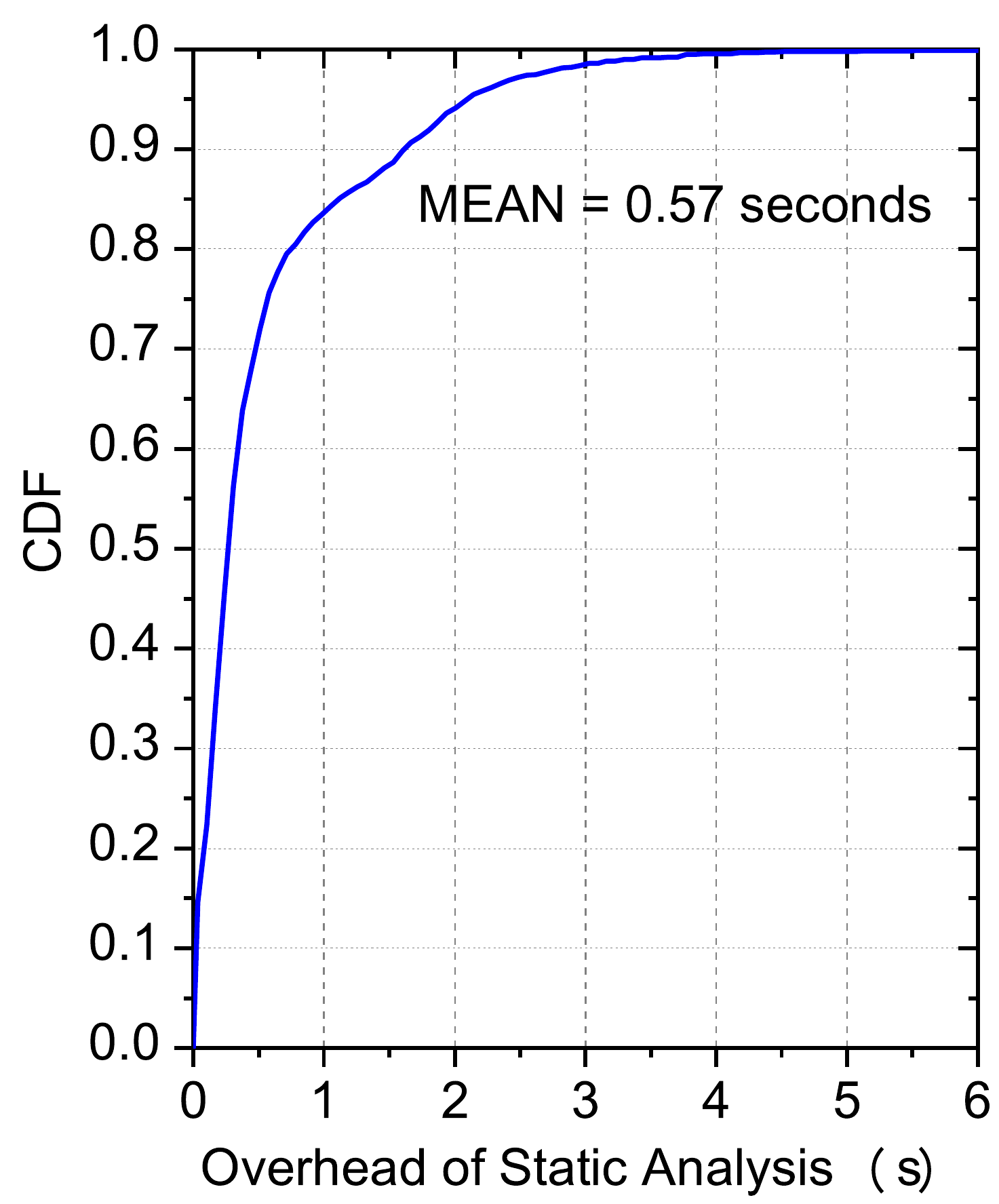}
\end{minipage}
}
\subfigure{
\begin{minipage}[t]{0.225\textwidth}
\centering
\includegraphics[width=\textwidth]{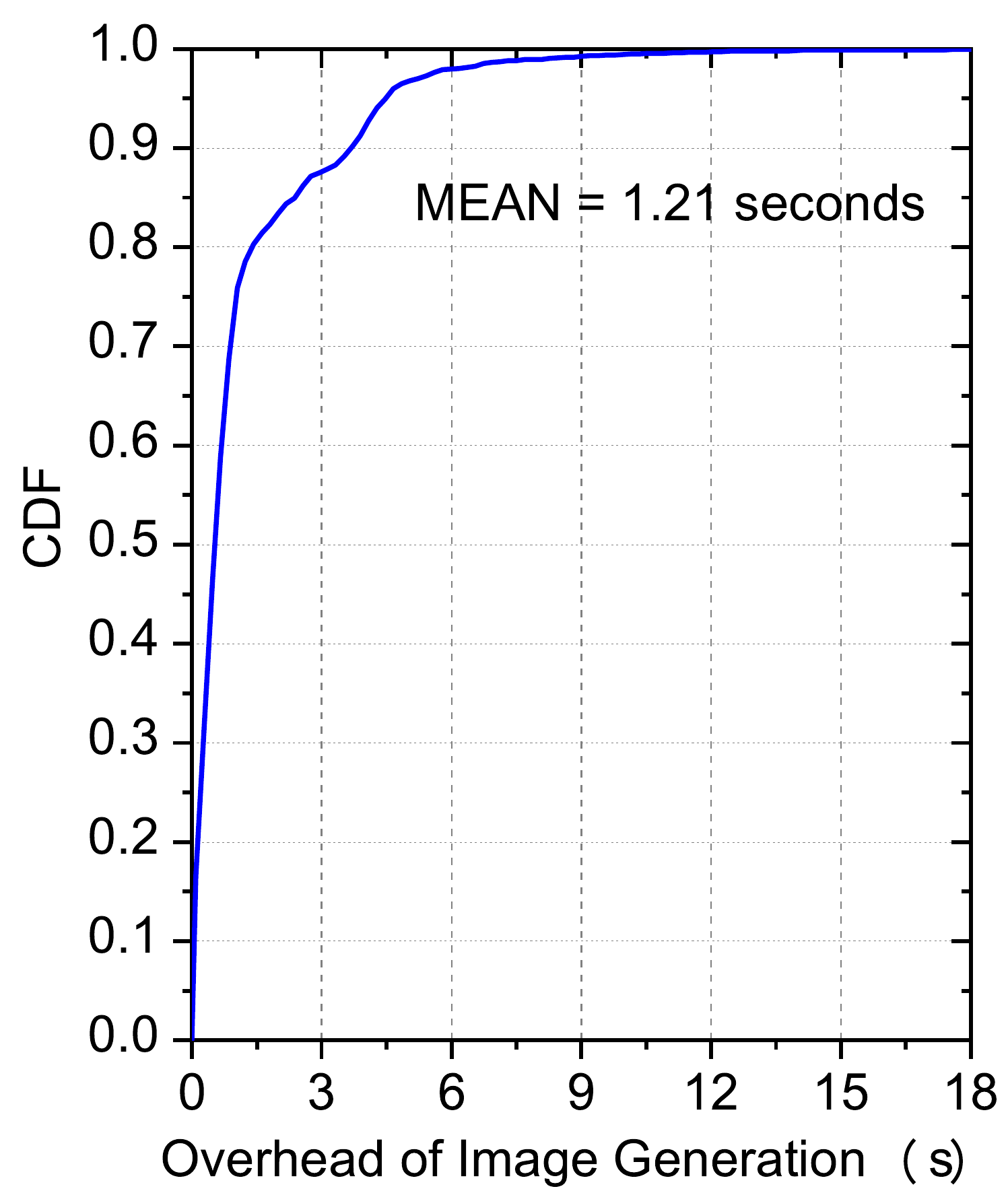}
\end{minipage}
}
\subfigure{
\begin{minipage}[t]{0.22\textwidth}
\centering
\includegraphics[width=\textwidth]{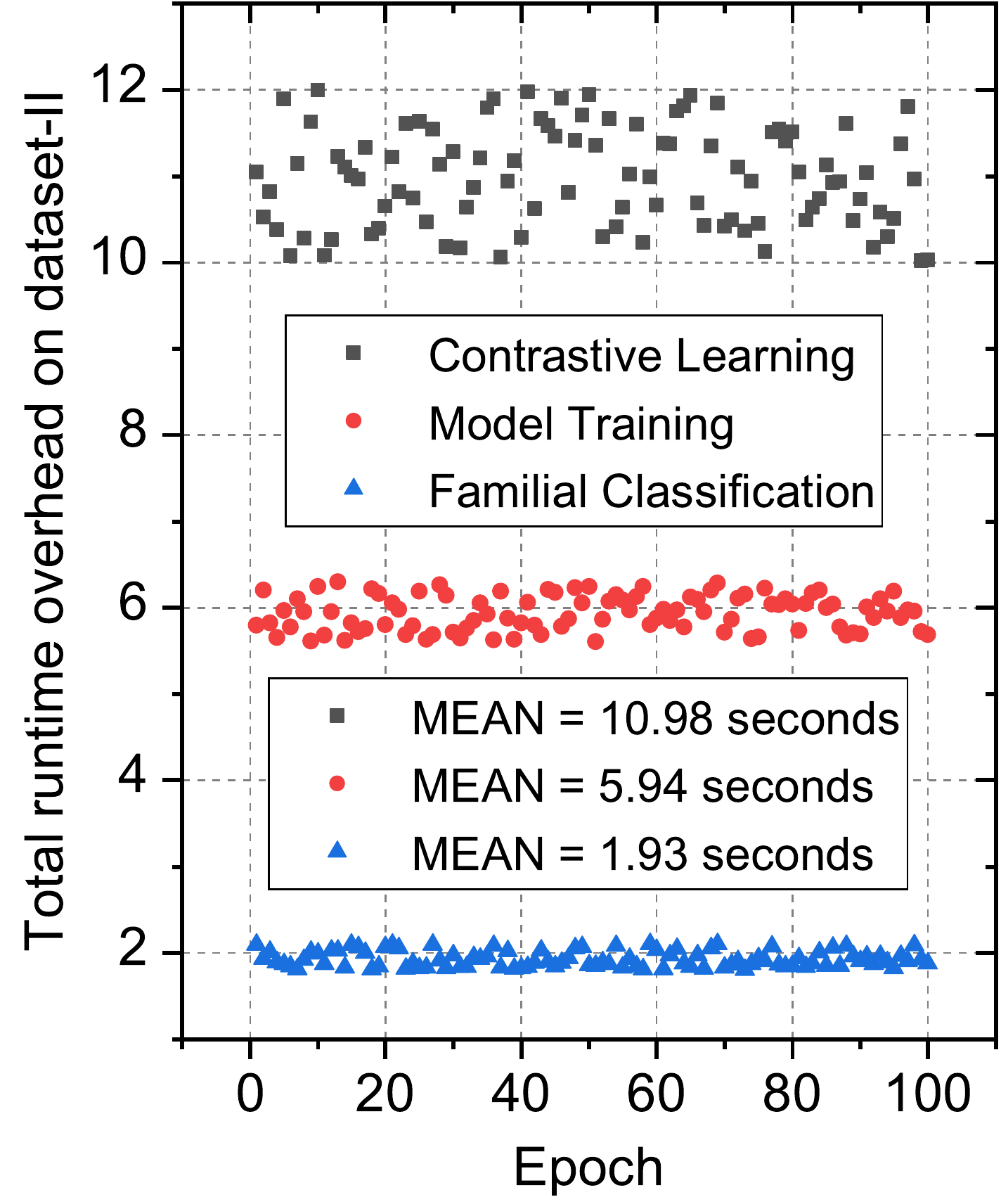}
\end{minipage}
}
\subfigure{
\begin{minipage}[t]{0.232\textwidth}
\centering
\includegraphics[width=\textwidth]{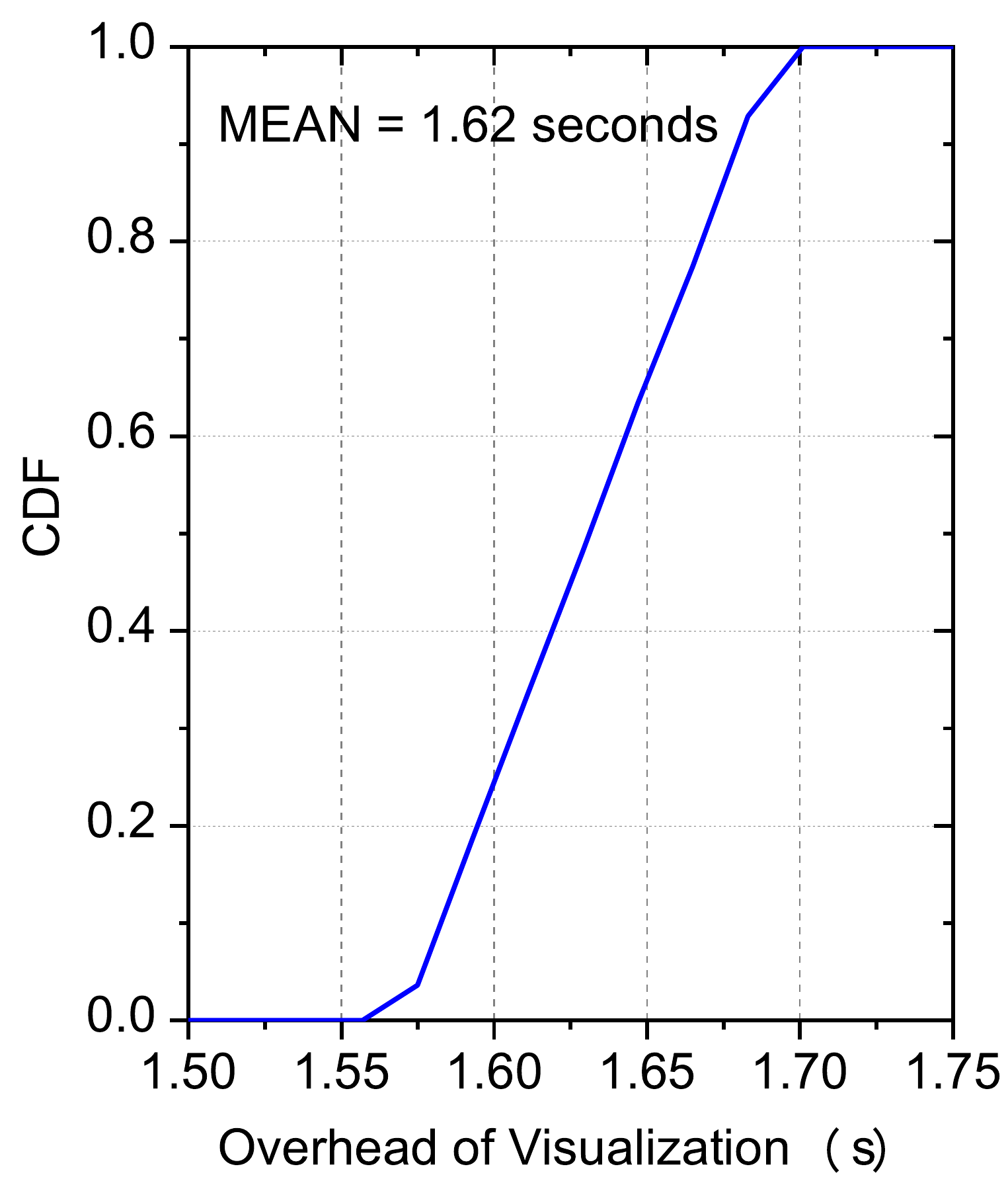}
\end{minipage}
}
\caption{\emph{Cumulative Distribution Function} (CDF) of runtime overheads of \emph{IFDroid} on different steps (seconds)}
\label{fig:overhead}
\end{figure}

In this step, we aim to study the runtime overhead of \emph{IFDroid}.
To this end, we randomly select 8,000 samples from dataset-II as our test target.
Given a new malware sample, \emph{IFDroid} performs four steps to classify it into the corresponding family and interpret the classification result.

\emph{1) Static Analysis.}
The first step of \emph{IFDroid} is to distill the program semantics of a sample into a function call graph based on static analysis.
Figure \ref{fig:overhead} and Table \ref{tab:overhead} show the runtime overhead of static analysis on dataset-II, for more than 80\% samples we can extract the graphs in one second.
On average, it takes about 0.57 seconds for \emph{IFDroid} to complete static analysis on dataset-II.

\emph{2) Image Generation.}
The second step of \emph{IFDroid} is to transform the call graph into an image to avoid high-cost graph matching.
Specifically, we extract four different centralities (\ie degree centrality, katz centrality, closeness centrality, and harmonic centrality) of sensitive API calls within the graph to construct the image.
As shown in Figure \ref{fig:overhead} and Table \ref{tab:overhead}, \emph{IFDroid} requires about 1.21 seconds on average to analyze the graph and complete the image generation step.

\emph{3) Familial Classification.}
The third step of \emph{IFDroid} is to classify the image into its corresponding family.
We first perform contrastive learning to learn an encoder and then train a classifier on generated images. 
The total runtime overhead of contrastive learning and model training on these images are shown in Figure \ref{fig:overhead}.
On average, it takes about 10.98 seconds and 5.94 seconds to finish a round of contrastive learning and model training, respectively.
After completing the training phase, we leverage the learned encoder and the trained classifier to complete the familial classification step of \emph{IFDroid}.
As shown in Figure \ref{fig:overhead} and Table \ref{tab:overhead}, this step is the fastest step among all step in \emph{IFDroid}.
It consumes about 1.93 seconds to classify all images (\ie 8,000 images) into corresponding families. 

\emph{4) Interpretation.}
The final step of \emph{IFDroid} is to interpret the classification results of 8,000 images.
To this end, we use Grad-CAM++ visualization technique to obtain the corresponding heatmaps of these images.
This step is the most time-consuming, it requires 1.62 seconds on average to accomplish the visualization of an image.

\begin{table}[htbp]
\footnotesize
  \centering
  \caption{Runtime overhead of different steps of \emph{IFDroid} on 8,000 malware samples}
    \begin{tabular}{|c|c|c|}
    \hline
    Different Steps & Total runtime & Average runtime \\
    \hline
    Static Analysis & 4,792 & 0.57 \\
    Image Generation & 10,172 & 1.21 \\
    Familial Classification & 1.93  & 0.00023 \\
    Interpretation & 13,619 & 1.62 \\
    \hline
    ALL   & 28,585 & 1.78+1.62=3.4 \\
    \hline
    \end{tabular}%
  \label{tab:overhead}%
\end{table}%

In general, given a new malware sample, \emph{IFDroid} consumes about 1.78 seconds to complete the classification and 1.62 seconds to interpret the classification result.
As for \emph{FalDroid}, \emph{AOM}, \emph{MVIIDroid}, and \emph{CDFG}, they need to take about 11.5 seconds, 9.2 seconds, 12.4 seconds, and 26.9 seconds to complete the classification of a malware sample.
In other words, if only from the overhead caused by classification, \emph{IFDroid} is about six times, five times, seven times, and 15 times faster than \emph{FalDroid}, \emph{AOM}, \emph{MVIIDroid}, and \emph{CDFG}.
Such high efficiency indicates that \emph{IFDroid} can achieve large-scale Android malware classification and interpretation.

\emph{\textbf{Summary:} On average, IFDroid requires about 1.78 seconds to complete the classification and 1.62 seconds to interpret the classification result of a malware sample.}

\section{Discussions}

\subsection{Differences from \emph{MalScan}}
In our previous work (\ie \emph{MalScan}~\cite{wu2019malscan}), we use centrality analysis to accomplish scalable Android malware detection.
The goal of \emph{MalScan} is to distinguish malware samples from benign apps while \emph{IFDroid} aims to classify malware into corresponding families.
In reality, we also apply \emph{MalScan} to classify our dataset-II, but the result is not ideal.
The classification accuracy on 15 families is only 85.2\%.
In other words, \emph{MalScan} is not suitable for Android malware familial classification.
To address the issue, we develop \emph{IFDroid} which is a robust and interpretable Android malware familial classification system.

\subsection{Why is IFDroid obfuscation-resilient?}
The reasons are mainly three-fold. 
First, \emph{IFDroid} uses sensitive API calls to form the features and API calls are not obfuscated.
Second, \emph{IFDroid} applies centrality analysis to maintain the graph details which is robust against obfuscations.
Last and most important, the learned encoder by contrastive learning in \emph{IFDroid} can extract robust features from generated images.
The goal of contrastive learning is to maximize the agreement between positive data and minimize the agreement between negative data.
Actually, the obfuscated malware can be regarded as one positive sample of the original malware since the inherent program semantics do not change after obfuscations.
Therefore, when we use contrastive learning to learn the encoder, it can enlarge the similarity between obfuscated malware and original malware, making it possible to classify the obfuscated malware into its correct family.

\subsection{The Reasons of Image Generation}
The reasons are mainly three-fold.
First, CNN can support large-scale image analysis.
If we can transform a malware sample into an image, then we can use the CNN model to achieve large-scale malware analysis.
Second, convolution kernels of different sizes in CNN can automatically extract features from images.
In our generated image, each pixel represents a certain centrality value for a sensitive API call.
When different sizes of convolution kernels are used, different centrality values of different sensitive API calls can be combined to find the most suitable combined features for malware classification.
Third, CNN can be interpreted by some visualization techniques.
If we can employ some suitable visualization techniques, then we can interpret the results of malware family classification, making it clear to the security researcher why the sample is classified into this family.

\subsection{The Selection of Static Analysis}
Compared with dynamic analysis, static analysis does have some shortcomings. 
However, since different events need to be generated to trigger different behaviors, it is very expensive to dynamically analyze an app. 
Sometimes it may take hours to analyze an app. 
Such a huge overhead makes it unsuitable for large-scale malware analysis. 
In the real world, according to the AV-TEST Institute report \cite{AVTEST}, an average of about 9,000 Android malware samples were detected every day in 2021. 
If we use dynamic analysis to analyze the families of these malware samples, it is difficult to achieve daily malware scanning. But the average runtime overhead of our proposed method \emph{IFDroid} is only 3.4 seconds.
In other words, if we analyze 16 malware samples at a time, we only need about half an hour to complete all the analysis and interpretation of detection results. 
Moreover, according to the results in our paper, \emph{IFDroid} can achieve ideal performance in analyzing obfuscated malware due to the use of contrastive learning.

\subsection{Limitations}
\subsubsection{Call Graph}
To achieve efficient static analysis, we leverage Androguard \cite{desnos2011androguard} to extract the function call graph of a malware sample.
This graph is a context- and flow-insensitive call graph.
Moreover, malware samples can use reflection \cite{rastogi2013catch} to call sensitive API calls, in this case, we may miss the call relationships between these methods.
As shown in Table \ref{tab:obfuscation}, the \emph{CallIndirection} obfuscator significantly lowers the scores of all classifiers.
The F1 of \emph{IFDroid} with contrastive learning is only 91.1\% which is not good enough to distinguish these obfuscated malware.
In our future work, we plan to use advanced program analysis~\cite{li2016droidra} to generate more accurate call graphs to maintain better robustness against obfuscation.

\subsubsection{Sensitive API calls}
Sensitive API calls used in \emph{IFDroid} consist of 426 API calls that are highly correlated with malicious operations \cite{Liangyi2020Experiences}.
They occupy a small part of the whole sensitive API calls. 
We plan to conduct statistical analysis to select more valuable sensitive API calls and use them to generate our images.

As API calls invoked by Android malware may evolve over time, \emph{IFDroid} may suffer from the issue.
To resist false positives and false negatives caused by the evolution of Android malware, we will update our sensitive API calls in real-time and adopt the technique in \cite{zhang2020enhancing} to improve our classification effectiveness.

\subsubsection{Image Size}
After extracting four centrality values of 426 sensitive API calls, we can obtain a 426$*$4 vector representation.
To make our interpretation more intuitive, we crop the vector and turn it into a square image.
Specifically, we add 60 zeros at the end and then reshape it as a 42$*$42 (\ie 426$*$4+60=42$*$42) vector.
Although every pixel in our image is meaningful, the centrality values of the same API call may be segmented adjacent to the beginning and end of different lines.
Such case may affect the learning effectiveness of our model.
In our future work, we plan to select different image size to commence our experiments.
By this, we can find a more suitable size and achieve more effective results. 
In addition, we will also build a separate image for each centrality and combine the four images to analyze Android malware.

\subsubsection{Model}
In \emph{MalScan}, we have shown that centrality is not robust enough against tailored adversarial attacks.
In other words, an attacker might adjust the frequency of some sensitive API calls by adding some dead code, so that the model incorrectly classifies samples into another family.
However, we choose four different centralities, each of which can represent a kind of structural information of the call graph.
For example, degree centrality considers the degree of all nodes within a network, while closeness centrality analyzes the average shortest distance of all nodes.
Therefore, when an adversary attacks our model, he may consider four different centrality extraction algorithms to craft adversarial examples, which may result in a large attack overhead. 
In the future, we plan to choose more different centralities to preserve more graph details of the call graph, making it more robust against adversarial attacks.

When faced with a new family of malware samples, we cannot use the original model to classify them since the model has not seen the family.
However, instead of retraining the model, we can use a single-task continuous learning \cite{hadsell2020embracing} approach to mitigate the limitation.
Based on the original model, we only need to use a small number of new family samples to iterate the model in some steps, and then the model can flag the new family.

\section{Related Work}

\subsection{Malware Familial Classification}
Recently, many studies~\cite{suarez2014dendroid, feng2014apposcopy, zhang2014droidsift, avdiienko2015mudflow, deshotels2014droidlegacy, feng2016astroid, fan2018android, meng2016smart, cai2018droidcat, yang2014droidminer, garcia2018lightweight, wu2020mviidroidMVI, blanc2019identifyingAOM, zhiwu2019androidCDFG, feng2020performance, feng2020seqmobile} have been proposed to classify malware samples into corresponding families.
For example, \emph{Dendroid}~\cite{suarez2014dendroid} applies text mining techniques to analyze the code structures of Android malware and classify them into corresponding families.
\emph{Apposcopy}~\cite{feng2014apposcopy} considers both data-flow and control-flow information of malware samples to classify them by performing heavy-weight program analysis.
\emph{MudFlow}~\cite{avdiienko2015mudflow} extracts the source-and-sink pairs of malware samples and regards them as features to classify malware. 
\emph{FalDroid}~\cite{fan2018android} conducts frequent subgraph analysis to extract common subgraphs of each family and uses them to perform familial classification.
\emph{DroidSIFT}~\cite{zhang2014droidsift} extracts the weighted contextual API dependency graph to solve the malware deformation problem based on static analysis.
These proposed approaches consider different program information to achieve accurate malware classification.
However, heavy-weight program analysis results in low scalability, making them can not scale to large numbers of malware analysis.
Moreover, most of them only provide the corresponding labels (\ie families) to users and can not interpret the classification results.

\subsection{Contrastive Learning}
Contrastive learning was first introduced by Mikolov \emph{et al.}~\cite{2013Distributed} in 2013 for \emph{natural language processing} (NLP).
In recent years, it has been more and more popular on different NLP tasks such as text representation learning~\cite{2020DeCLUTR}, language understanding~\cite{2020Cert}, and cross-lingual pre-training~\cite{2020InfoXLM}.
In practice, it has also been used in other domains.
For example, Dai \emph{et al.}~\cite{2017Contrastive} propose a new method for image caption through contrastive learning. 
\emph{SimCLR}~\cite{2020A} is a simple framework to use contrastive learning on image classification.
Compared with previous work, the accuracy of \emph{SimCLR} is improved by 7\%.
\emph{COLA}~\cite{2020Contrastive} is a self-supervised pre-training approach for learning a general-purpose representation of audio and \emph{CVRL}~\cite{2020Spatiotemporal} uses contrastive learning to learn spatiotemporal visual representations from unlabeled videos.
The use of contrastive learning in most of the previous studies is self-supervised, however, Khosla \emph{et al.} \cite{khosla2020supervised} find that the label information of training dataset can improve the performance of the learned encoder.
Therefore, in this paper, to achieve better classification accuracy, we leverage supervised contrastive learning to conduct Android malware familial classification.

\section{Conclusion}

In this paper, we propose to use contrastive learning to resist code obfuscations of Android malware.
To demonstrate the effectiveness of contrastive learning, we implement an obfuscation-resilient system (\ie \emph{IFDroid}) and the extensive evaluation results show that \emph{IFDroid} is superior to ten state-of-the-art Android malware classification systems (\ie \emph{Dendroid}~\cite{suarez2014dendroid}, \emph{Apposcopy}~\cite{feng2014apposcopy}, \emph{DroidSIFT}~\cite{zhang2014droidsift}, \emph{MudFlow}~\cite{avdiienko2015mudflow}, \emph{DroidLegacy}~\cite{deshotels2014droidlegacy}, \emph{Astroid}~\cite{feng2016astroid}, \emph{FalDroid}~\cite{fan2018android}, \emph{AOM}~\cite{blanc2019identifyingAOM}, \emph{MVIIDroid}~\cite{wu2020mviidroidMVI}, and \emph{CDFG}~\cite{zhiwu2019androidCDFG}).
Moreover, when analyzing 69,421 obfuscated malware samples, \emph{IFDroid} can achieve a 98.4\% F1 score.
Such result also suggests that \emph{IFDroid} can achieve obfuscation-resilient malware analysis.
\ifCLASSOPTIONcompsoc
  \section*{Acknowledgments}
\else
  \section*{Acknowledgment}
\fi

This work is supported by the Key Program of National Science Foundation of China under Grant No. U1936211 and Hubei Province Key R\&D Technology Special Innovation Project under Grant No. 2021BAA032.

\ifCLASSOPTIONcaptionsoff
  \newpage
\fi

\bibliography{IFDroid}
\bibliographystyle{IEEEtran}

\begin{IEEEbiography}[{\includegraphics[width=1in,height=1.25in,clip,keepaspectratio]{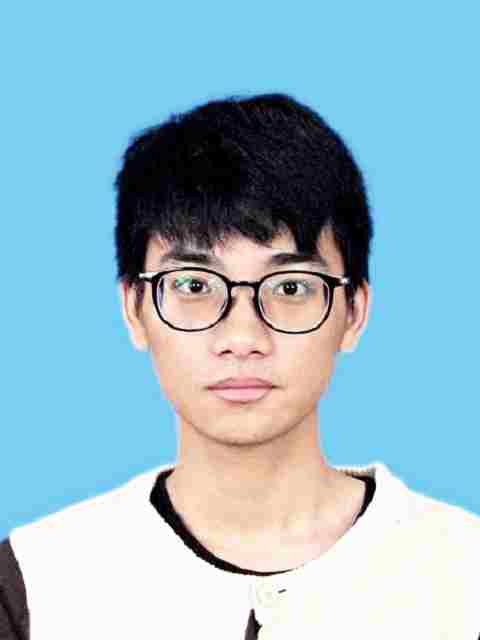}}]{Yueming Wu} received the B.E. degree in Computer Science and Technology at Southwest Jiaotong University, Chengdu, China, in 2016 and the Ph.D. degree in School of Cyber Science and Engineering at Huazhong University of Science and Technology, Wuhan, China, in 2021. He is currently a research fellow in the School of Computer Science and Engineering at Nanyang Technological University.
His primary research interests lie in malware analysis and vulnerability analysis.
\end{IEEEbiography}

\begin{IEEEbiography}[{\includegraphics[width=1in,height=1.25in,clip,keepaspectratio]{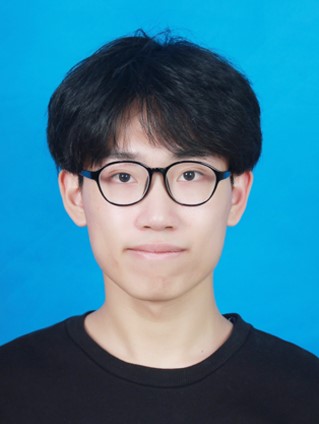}}]{Shihan Dou} received the B.E. degree in Cyberspace Security from Huazhong University of Science and technology, Wuhan, China, in 2021, and is currently pursuing the M.E. degree in School of Computer Science at Fudan University. His primary research interests lie in malware analysis and vulnerability analysis.
\end{IEEEbiography}

\begin{IEEEbiography}[{\includegraphics[width=1in,height=1.25in,clip,keepaspectratio]{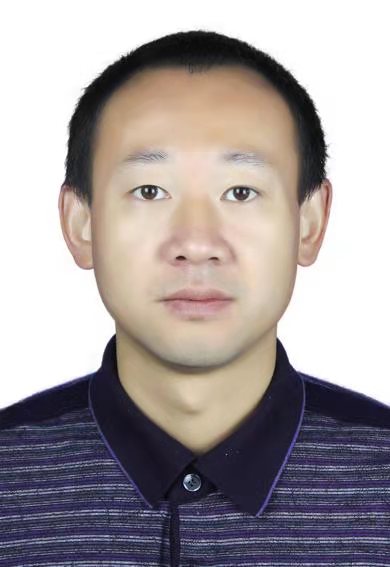}}]{Deqing Zou} received the Ph.D. degree at Huazhong University of Science and Technology (HUST), in 2004. He is currently a professor of School of Cyber Science and Engineering, Huazhong University of Science and Technology (HUST), Wuhan, China. His main research interests include system security, trusted computing, virtualization and cloud security. He has always served as a reviewer for several prestigious journals, such as IEEE TDSC, IEEE TOC, IEEE TPDS, and IEEE TCC. He is on the editorial boards of four international journals, and has served as PC chair/PC member of more than 40 international conferences.
\end{IEEEbiography}

\begin{IEEEbiography}[{\includegraphics[width=1in,height=1.25in,clip,keepaspectratio]{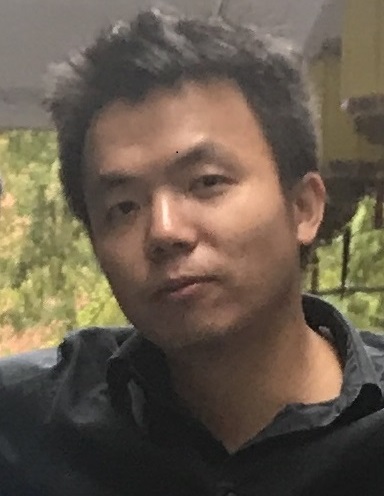}}]{Wei Yang} received his Ph.D. in Computer Science from the University of Illinois at Urbana-Champaign and his bachelor degree from Shanghai Jiao Tong University. He was a visiting researcher in University of California, Berkeley.
His research interests are in software engineering and security. His current primary projects relate to Mobile Security, Software engineering/Security for Machine Learning, Intelligent SE/Security, and IoT/Blockchain Security.
\end{IEEEbiography}

\begin{IEEEbiography}[{\includegraphics[width=1in,height=1.25in,clip,keepaspectratio]{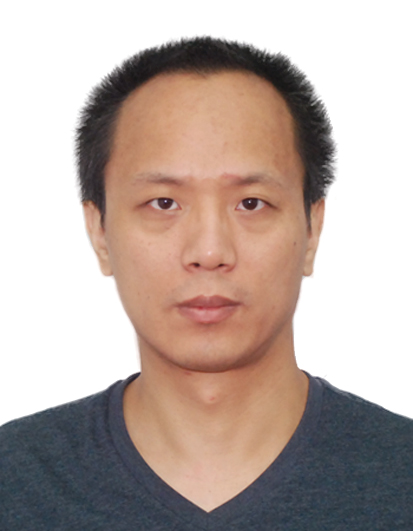}}]{Weizhong Qiang} received the PhD degree from Huazhong University of Science and Technology (HUST), in 2005. He is currently a professor of School of Cyber Science and Engineering, Huazhong University of Science and Technology (HUST), Wuhan, China. His research interest is mainly about software system security. He has published more than 30 research papers.
\end{IEEEbiography}

\begin{IEEEbiography}[{\includegraphics[width=1in,height=1.25in,clip,keepaspectratio]{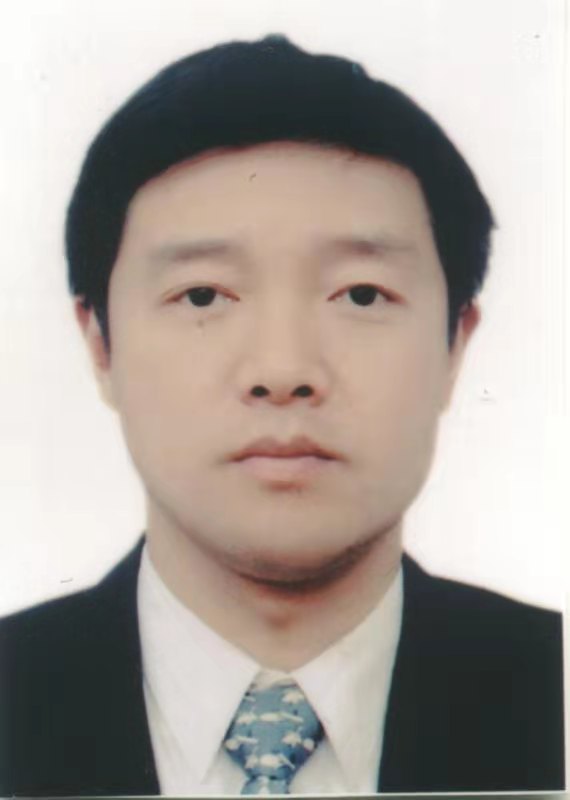}}]{Hai Jin} received the Ph.D. degree in computer engineering from Huazhong University of Science and Technology (HUST), Wuhan, China, in 1994. He is a Cheung Kung Scholars Chair Professor of computer science and engineering at HUST in China. He was awarded Excellent
Youth Award from the National Science Foundation
of China in 2001. He is the chief scientist of ChinaGrid, the largest grid computing project in
China, and the chief scientists of National 973
Basic Research Program Project of Virtualization
Technology of Computing System, and Cloud Security. He is a fellow of the IEEE, a fellow of the CCF, and a member of the ACM. He has co-authored 22 books and published over 700 research papers. His research interests include computer architecture, virtualization technology, cluster computing and cloud computing, peer-to-peer computing, network storage, and network security.
\end{IEEEbiography}

\end{document}